\documentclass[preprint,showpacs,amsmath,amssymb,aps,prd,nofootinbib]{revtex4}
\usepackage{epsfig,graphicx,color,appendix}
\begin{document}
\begin{flushright}
KIAS-P12008
\end{flushright}
\begin{center}
{\large \bf Non-zero $\theta_{13}$ linking to Dark Matter from
Non-Abelian Discrete Flavor Model in Radiative Seesaw}
\end{center}

\author{Y. H. Ahn\footnote{Email: yhahn@kias.re.kr}}
\author{Hiroshi Okada\footnote{Email: hokada@kias.re.kr}}

\affiliation{School of Physics, KIAS, Seoul 130-722, Korea}

\date{\today}

\begin{abstract}
We propose a new scenario in a radiative seesaw model based on $A_4$ flavor symmetry.
In this model, we explore a possibility of linking non-zero $\theta_{13}$ to dark matter.
And we analyze the lepton sector to predict the observed neutrinos and mixings, especially obtaining a lower bound of $\theta_{13}\gtrsim3.5^{\circ}$.
We show that the non-zero $\theta_{13}$ is
correlated with our heavy Majorana type of dark matter.
Also we predict that the mass be ${\cal O}$(1-10) TeV, as a result of analyzing the Wilkinson-Microwave-Anisotropy-Probe and lepton flavor violation.
\end{abstract}

\maketitle %
\section{Introduction}
The large values of the solar ($\theta_{12}$) and atmospheric ($\theta_{23}$) mixing angles in the Pontecorvo-Maki-Nakagawa-Sakata (PMNS) matrix \cite{pmns} may be telling us about some new symmetries of leptons not presented in the quark sector and may provide a clue to the nature of the quark-lepton physics beyond the standard model (SM). If there exists such a flavor symmetry in Nature, the tribimaximal (TBM)~\cite{HPS} pattern, $\sin^{2}\theta_{12}=1/3$, $\sin^{2}\theta_{23}=1/2$, $\sin\theta_{13}=0$, for the neutrino mixing will be a good zeroth order approximation to reality.
A Non-Abelian discrete symmetry \cite{ishi-al} could play an important role in predicting the mixing angles. For example, in a well-motivated extension of the SM through the inclusion of $A_{4}$ discrete symmetry, the TBM pattern comes out in a natural way in the work of~\cite{Ma:2001dn, Altarelli:2005yp, He:2006dk}. Although such a flavor symmetry is realized in Nature leading to exact TBM, in general there may be some deviations from TBM~\cite{Ahn:2011yj, Ahns}.
Recent data of the T2K, MINOS and Double Chooz~\cite{Data} Collaborations imply that the unknown mixing angle, $\theta_{13}$, can be relatively large, indicating a deviation from the exact TBM which leads to vanishing $CP$ violation in neutrino oscillations. And, the analysis based on global fits~\cite{GonzalezGarcia:2010er, Fogli:2011qn, valle} of neutrino oscillations enter into a new phase of precise measurements of the neutrino mixing angles and mass-squared differences, indicating that the TBM mixing for three flavors of leptons should be modified.
Their current best-fit values in $1\sigma$ $(3\sigma)$ ranges from neutrino oscillation experiments are given by~\cite{valle}
 \begin{eqnarray}
  &&\theta_{12}=34.0^{\circ+1.0^{\circ}~(+2.9^{\circ})}_{~-0.9^{\circ}~(-2.7^{\circ})}~,
  ~\quad\theta_{23}=46.1^{\circ+3.5^{\circ}~(+7.0^{\circ})}_{~-4.0^{\circ}~(-7.5^{\circ})}~,
  ~\quad\theta_{13}=6.5^{\circ+1.6^{\circ}~(+4.2^{\circ})}_{~-1.4^{\circ}~(-4.7^{\circ})}~,\nonumber\\
  && \Delta m^{2}_{21}[10^{-5}{\rm eV}^{2}]=7.59^{+0.20~(+0.60)}_{-0.18~(-0.50)}~,\quad\qquad\Delta m^{2}_{31}[10^{-3}{\rm eV}^{2}]=2.50^{+0.09~(+0.26)}_{-0.16~(-0.36)}~,
 \label{expnu}
 \end{eqnarray}
which corresponds to normal neutrino mass ordering~\footnote{See Ref.~\cite{valle} for the data of inverted mass ordering.}.
And we know nothing about all three $CP$-violating phases $\delta_{CP},~\delta_{1}$ and $\delta_{2}$.
Besides the mystery of the mixing pattern, tiny neutrino mass is one of the most challenging problem beyond SM. It suggests that neutrinos could be induced by a radiative correction or tiny couplings if a theory be within TeV scale. Several years ago, Ernest Ma introduced the so-called radiative seesaw mechanism~\cite{ma-rad} where the masses are generated through one-loop effects \cite{others}. Moreover, since the existence of the flavor neutrino mixing for the three neutrinos $\nu_e,\nu_\mu,\nu_\tau$ implies that the individual lepton charges, $L_\alpha,(\alpha=e,\mu,\tau)$, are not conserved~\cite{Bilenky:1987ty}, the observation of neutrino oscillation have the possibility of measurable branching ratio for charged lepton flavor violation (LFV) decays such as $\mu\rightarrow e\gamma,\tau\rightarrow e\gamma$ and $\tau\rightarrow\mu\gamma$, etc.. Experimental discovery of such lepton rare decay processes is one of smoking gun signals of physics beyond SM.

Our starting point is an effective Lagrangian obeying the discrete $A_{4}\times Z_{2}$ symmetry which is spontaneously broken by the vacuum expectation value (VEV) of $SU(2)_{L}\times U(1)_{Y}$ singlet scalar fields at a scale higher than the electroweak scale. In addition, we assign a $Z_{2}$-odd quantum number to a new Higgs doublet and three right-handed singlet fermions while all SM particles are $Z_{2}$-even parity, in order to explain both the TBM at tree level Lagrangian and dark matter (DM). After electroweak symmetry breaking, the $Z_{2}$ symmetry is exactly conserved and a $Z_{2}$-odd doublet Higgs that does not have VEV, so called {\it inert Higgs}, while the standard Higgs boson get a VEV, which means the Yukawa coupling corresponding to $Z_{2}$-odd Higgs doublet will not generate the Dirac mass terms in neutrino sector. Thus, the usual seesaw mechanism does not work any more and we naturally have a good candidate of DM corresponding to the lightest $Z_{2}$-odd particle or Large Hadron Collider (LHC) signals through the standard gauge interactions in our scenario. The assigned leptonic flavor symmetry will lead us to a neutrino mass matrix through one-loop mediated by a new Higgs doublet and right-handed neutrinos having $Z_{2}$-odd parity, and indicating the deviations from TBM in lepton sector by dimension-5 effective operators driven by $SU(2)_{L}\times U(1)_{Y}$ singlet scalar fields with a cutoff scale $\Lambda$. Here we assume that $CP$ is a good symmetry above the cut-off scale $\Lambda$ in Lagrangian level. In this paper, we address the possibility of a linking between DM and a non-zero $\theta_{13}$ through the combination of WMAP results~\cite{wmap} with LFV $\mu\to e\gamma$ decay. We analyze possible spectrums of light neutrinos and their flavor mixing angles. In order to investigate the relation between DM and a non-zero $\theta_{13}$, we focus on a normal hierarchical mass spectrum of light neutrinos.

This paper is organized as follows. In section II, we show our model building for the lepton sector, in which we adopt $A_4$ group.
In section III, we discuss the predictions for neutrinos coming from the flavor symmetry. In section IV, we show constraints of DM mass from WMAP and LFV;
$\mu \to e\gamma$, $\tau \to e\gamma$, and $\tau \to \mu\gamma$ processes.
Section V is devoted to conclusions/discussions. We discuss the Higgs potential in the appendix.
\section{flavor $A_{4}$ symmetry and a discrete symmetry $Z_{2}$}
In the absence of flavor symmetries, particle masses and mixings
are generally undetermined in gauge theory. We work in the
framework of the SM, extended to consist of the
right-handed $SU(2)_{L}$-singlet Majorana neutrinos, $N_{R}$. The
scalar sector, apart from the usual SM Higgs doublet $\Phi$, is
extended through the introduction of two types of scalar fields,
$\chi$ and $\eta$, that are singlet and doublet under
$SU(2)_{L}\times U(1)_{Y}$, respectively:
 \begin{eqnarray}
  \Phi =
  {\left(
  \varphi^{+}~
  \varphi^{0}
\right)^{T}}
~,~~~\chi~,~~~\eta =
  {\left(
  \eta^{+}~
  \eta^{0}
 \right)^{T}}~.
  \label{Higgs}
 \end{eqnarray}
Furthermore, to understand the present neutrino oscillation data we consider $A_{4}$ flavor symmetry for leptons, and simultaneously for both the existence of DM and TBM at tree-level Lagrangian to be explained we also introduce an auxiliary discrete symmetry $Z_{2}$ in a radiative seesaw~\cite{ma-rad}.
Here we recall that $A_{4}$ is the symmetry group of the
tetrahedron and the finite groups of the even permutation of four
objects. Its irreducible representations contain one triplet ${\bf
3}$ and three singlets ${\bf 1}, {\bf 1}', {\bf 1}''$ with the
multiplication rules ${\bf 3}\otimes{\bf 3}={\bf 3}_{s}\oplus{\bf
3}_{a}\oplus{\bf 1}\oplus{\bf 1}'\oplus{\bf 1}''$, ${\bf
1}'\otimes{\bf 1}''={\bf 1}$, ${\bf 1}'\otimes{\bf 1}'={\bf 1}''$
and ${\bf 1}''\otimes{\bf 1}''={\bf 1}'$. Let us denote $(a_{1},
a_{2}, a_{3})$ and $(b_{1}, b_{2}, b_{3})$ as two $A_4$ triplets,
then we have
 \begin{eqnarray}
  (a\otimes b)_{{\bf 3}_{\rm s}} &=& (a_{2}b_{3}+a_{3}b_{2}, a_{3}b_{1}+a_{1}b_{3}, a_{1}b_{2}+a_{2}b_{1})~,\nonumber\\
  (a\otimes b)_{{\bf 3}_{\rm a}} &=& (a_{2}b_{3}-a_{3}b_{2}, a_{3}b_{1}-a_{1}b_{3}, a_{1}b_{2}-a_{2}b_{1})~,\nonumber\\
  (a\otimes b)_{{\bf 1}} &=& a_{1}b_{1}+a_{2}b_{2}+a_{3}b_{3}~,\nonumber\\
  (a\otimes b)_{{\bf 1}'} &=& a_{1}b_{1}+\omega a_{2}b_{2}+\omega^{2}a_{3}b_{3}~,\nonumber\\
  (a\otimes b)_{{\bf 1}''} &=& a_{1}b_{1}+\omega^{2} a_{2}b_{2}+\omega a_{3}b_{3}~,
 \end{eqnarray}
where $\omega=e^{i2\pi/3}$ is a complex cubic-root of unity.

The field contents under $SU(2)\times U(1)\times A_{4}\times
Z_{2}$ of the model are assigned in Table~\ref{reps} :
\begin{widetext}
\begin{center}
\begin{table}[h]
\caption{\label{reps} Representations of the fields under $A_4\times Z_{2}$ and $SU(2)_L \times U(1)_Y$.}
\begin{ruledtabular}
\begin{tabular}{ccccccccccccc}
Field &$L_{L}$&$l_R,l'_R,l''_R$&$N_{R}$&$\chi$&$\Phi$&$\eta$\\
\hline
$A_4$&$\mathbf{3}$&$\mathbf{1}$, $\mathbf{1^\prime}$,$\mathbf{1^{\prime\prime}}$&$\mathbf{3}$&$\mathbf{3}$&$\mathbf{3}$&$\mathbf{1}$\\
$Z_2$&$+$&$+$&$-$&$+$&$+$&$-$\\
$SU(2)_L\times U(1)_Y$&$(2,-1)$&$(1,-2)$&$(1,0)$&$(1,0)$&$(2,1)$&$(2,1)$\\
\end{tabular}
\end{ruledtabular}
\end{table}
\end{center}
\end{widetext}
At the Lagrangian level, we shall assume the absence of CP violation in the Dirac neutrino sector and in charged lepton Yukawa interactions above the cutoff scale $\Lambda$, which for scales below $\Lambda$ is expressed in terms of effective dimension-5 operators.
With dimension-5 operators driven by $\chi$ fields the Yukawa interactions $(d\leq5)$ in the neutrino and charged lepton sectors invariant under $SU(2)\times U(1)\times A_{4}\times Z_{2}$ can be written as
 \begin{eqnarray}
 -{\cal L}_{\rm Yuk} &=& y_{\nu}(\bar{L}_{L}N_{R})_{{\bf 1}}\tilde{\eta}
 +\frac{1}{2}M(\overline{N^{c}_{R}}N_{R})_{{\bf 1}}+\frac{1}{2}\lambda_{\chi}(\overline{N^{c}_{R}}N_{R})_{{\bf 3}_{s}} \chi\nonumber\\
 &+&\frac{y^{s}_{N}}{\Lambda}[(\bar{L}_{L}N_{R})_{{\bf 3}_{s}}\chi]\tilde{\eta}+\frac{y^{a}_{N}}{\Lambda}[(\bar{L}_{L}N_{R})_{{\bf 3}_{a}}\chi]\tilde{\eta}\nonumber\\
 &+& y_{e}(\bar{L}_{L}\Phi)_{{\bf 1}}l_{R}+y_{\mu}(\bar{L}_{L}\Phi)_{{\bf 1}'}l''_{R}+y_{\tau}(\bar{L}_{L}\Phi)_{{\bf 1}''}l'_{R}\nonumber\\
 &+&\frac{y^{s}_{e}}{\Lambda}[(\bar{L}_{L}\Phi)_{{\bf 3}_{s}}\chi]l_{R}+\frac{y^{s}_{\mu}}{\Lambda}[(\bar{L}_{L}\Phi)_{{\bf 3}_{s}}\chi]_{{\bf 1}'}l''_{R}+\frac{y^{s}_{\tau}}{\Lambda}[(\bar{L}_{L}\Phi)_{{\bf 3}_{s}}\chi]_{{\bf 1}''}l'_{R}\nonumber\\
 &+&\frac{y^{a}_{e}}{\Lambda}[(\bar{L}_{L}\Phi)_{{\bf 3}_{a}}\chi]l_{R}+\frac{y^{a}_{\mu}}{\Lambda}[(\bar{L}_{L}\Phi)_{{\bf 3}_{a}}\chi]_{{\bf 1}'}l''_{R}+\frac{y^{a}_{\tau}}{\Lambda}[(\bar{L}_{L}\Phi)_{{\bf 3}_{a}}\chi]_{{\bf 1}''}l'_{R}+h.c~,
 \label{lagrangian}
 \end{eqnarray}
where $\tilde{\eta}\equiv i\tau_{2}\eta^{\ast}$ and
$L_{L}=(\nu_{L},\ell^{-}_{L})^T$ are the Higgs doublet and the
lepton doublet transforming as singlet ${\bf 1}$ and triplet ${\bf
3}$ under $A_{4}$, respectively, and here $\tau_{2}$ is the Pauli
matrix. In the above Lagrangian, heavy neutrinos acqiure a bare
mass term $M$ and a mass induced by the electroweak singlet $\chi$
scalar with ${\bf 3}$ representation under $A_4$. By imposing an
additional symmetry $Z_{2}$ as shown in Table~\ref{reps}, the
$A_{4}\times SU(2)_{L}\times U(1)_{Y}$ invariant Yukawa term
$\bar{\ell}_{L}N_{R}\Phi$ is forbidden from the Lagrangian, and the neutral component of scalar doublet $\eta$ will not generate a VEV,
 \begin{eqnarray}
  \langle\eta^{0}\rangle\equiv\upsilon_{\eta}=0~.
 \end{eqnarray}
Therefore, the scalar field $\eta$ can only couple to the standard gauge bosons as well as the Dirac neutrino mass terms are vanished, which means the usual seesaw does not operate anymore. However, the light neutrino Majorana neutrio mass matrix can be generated radiatively through one-loop with the help of the Yukawa interaction $\bar{L}_{L}N_{R}\tilde{\eta}$ and $\bar{L}_{L}N_{R}\chi\tilde{\eta}/\Lambda$ in the Lagrangian, we will discuss it more detaily in Sec.III.

Taking the $A_{4}$ symmetry breaking scale above the electroweak
scale in our scenario, that is, $\langle\chi\rangle >
\langle\Phi^{0}\rangle$, and assuming the vacuum alignment of
fields $\langle\chi_{i}\rangle$ as
 \begin{eqnarray}
  \langle\chi_{1}\rangle&\equiv&\upsilon_{\chi}\neq0,~\langle\chi_{2}\rangle=\langle\chi_{3}\rangle=0~,
 \label{subgroup}
 \end{eqnarray}
 then the right-handed neutrino Majorana mass terms are given by
 \begin{eqnarray}
 M_{R}=M{\left(\begin{array}{ccc}
 1 &  0 &  0 \\
 0 &  1 &  \kappa e^{i\xi} \\
 0 &  \kappa e^{i\xi} &  1
 \end{array}\right)}~,
 \label{MR2}
 \end{eqnarray}
where $\kappa=|\lambda^{s}_{\chi}\upsilon_{\chi}/M|$ with $\langle\chi_{i}\rangle=\upsilon_{\chi_{i}}~(i=1,2,3)$. After the electroweak symmetry breaking $\langle\eta^{0}\rangle=\upsilon$ with the VEV alignment in Eq.~(\ref{subgroup}), the Yukawa interaction $y_{\nu}\bar{L}_{L}N_{R}\tilde{\eta}$, together with the terms $\frac{y^{s,a}_{N}}{\Lambda}[(\bar{L}_{L}N_{R})_{{\bf 3}_{s}}\chi]\cdot\tilde{\eta}$ can be written as $\upsilon\overline{\nu}_{L}Y_{\nu}N_{R}$ with the neutrino Yukawa coupling matrix $Y_{\nu}$ given by
 \begin{eqnarray}
 Y_{\nu}=e^{i\rho}|y_{\nu}|{\left(\begin{array}{ccc}
 1 &  0 &  0 \\
 0 & 1 & y_{1}e^{i\rho_{1}} \\
 0 &  y_{2}e^{i\rho_{2}} & 1
 \end{array}\right)}~,
 \label{yukawaNu}
 \end{eqnarray}
where $y_{1}=|y^{s}_{N}+y^{a}_{N}|\upsilon_{\chi}/|y_{\nu}|\Lambda$, $y_{2}=|y^{s}_{N}-y^{a}_{N}|\upsilon_{\chi}/|y_{\nu}|\Lambda$ and $\rho=\arg(y_{\nu})$. Eq.~(\ref{yukawaNu}) indicates that, once the VEV alignment in Eq.~(\ref{subgroup}) is taken, the $A_{4}$ symmetry is spontaneously broken and its residual symmetry $Z_{2}$~\cite{Ahn:2011yj, He:2006dk} is also explicitly broken by the higher dimensional operators. Therefore, we can expect a low energy CP violation responsible for the neutrino oscillation as well as high energy CP violation responsible for the leptogenesis in the neutrino sector, which could be generated by the off-diagonal terms of Yukawa neutrino coupling matrix~\cite{Ahn:2011yj, Ahn:2010cc}.

Assuming equally aligned VEVs of $A_{4}$ triplets, $\langle\Phi^{0}\rangle=(\upsilon,\upsilon,\upsilon)$, together with the VEV alignment in Eq.~(\ref{subgroup}), the light charged lepton mass matrix can be explicitly expressed as
 \begin{eqnarray}
 m_{\ell}&=&U_{\omega}\sqrt{3}{\left(\begin{array}{ccc}
 m^{\ell}_{11} & m^{\ell}_{12} & m^{\ell}_{13} \\
 m^{\ell}_{21} & m^{\ell}_{22} & m^{\ell}_{23} \\
 m^{\ell}_{31} & m^{\ell}_{32} & m^{\ell}_{33}
 \end{array}\right)}~,~~~~~~~{\rm with}~~
 U_{\omega}=\frac{1}{\sqrt{3}}{\left(\begin{array}{ccc}
 1 &  1 &  1 \\
 1 &  \omega &  \omega^{2} \\
 1 &  \omega^{2} &  \omega
 \end{array}\right)}\nonumber\\
 &=& U_{\omega}V^{\ell}_{L}{\rm Diag.}(m_{e},m_{\mu},m_{\tau})V^{\ell\dag}_{R}~,
 \label{CHcorrect}
 \end{eqnarray}
where $U_{\omega}V^{\ell}_{L}$ and $V^{\ell\dag}_{R}$ indicate that the left and right-diagonalization matrices for the charged lepton sector, and the mass matrix elements are written as
 \begin{eqnarray}
  m^{\ell}_{11}&=&\upsilon(y_{f_{1}}+2h_{1}/3) ,~~m^{\ell}_{12}=2\upsilon h_{2}/3,~~~~~~~~~~~~~~~~~~~m^{\ell}_{13}=2\upsilon h_{3}/3~, \nonumber\\ m^{\ell}_{21}&=&\upsilon(g_{1}-h_{1})/3 ,~~~~~m^{\ell}_{22}=\upsilon(y_{f_{2}}+(g_{2}-h_{2})/3),~~m^{\ell}_{23}=\upsilon(g_{3}-h_{3})/3~,\nonumber\\ m^{\ell}_{31}&=&-\upsilon(g_{1}+h_{1})/3 ,~~~m^{\ell}_{32}=-\upsilon(g_{2}+h_{2})/3,~~~~~~~~~m^{\ell}_{33}=\upsilon(y_{f_{3}}-(g_{3}+h_{3})/3)~,
 \end{eqnarray}
with $h_{1}=\upsilon_{\chi}y^{s}_{f_{1}}/\Lambda$,
$h_{2}=\upsilon_{\chi}y^{s}_{f_{2}}/\Lambda$,
$h_{3}=\upsilon_{\chi}y^{s}_{f_{3}}/\Lambda$,
$g_{1}=-i\sqrt{3}\upsilon_{\chi}y^{a}_{f_{1}}/\Lambda$,
$g_{2}=-i\sqrt{3}\upsilon_{\chi}y^{a}_{f_{2}}/\Lambda$,
$g_{3}=-i\sqrt{3}\upsilon_{\chi}y^{a}_{f_{3}}/\Lambda$. All of
them are complex. The important point is that, taking the VEV
alignment of $\langle\chi_{i}\rangle$ in Eq.~(\ref{subgroup})
together with the equal VEV alignment of $\langle\Phi^{0}\rangle$,
the $A_{4}$ symmetry is broken and its residual $C_{3}$
symmetry~\cite{He:2006dk} is also broken through the dimension-5
operators. One of the most striking features in the charged lepton
sector is that the mass spectra of charged leptons are strongly
hierarchical, {\it i.e.} the third generation fermions are much
heavier than the first and second generation fermions. From
Eq.~(\ref{CHcorrect}), for the most natural case of hierarchical
charged lepton Yukawa couplings $y_{f_{3}}\gg y_{f_{2}}\gg
y_{f_{1}}$, the corrected off-diagonal terms which are from
dimension-5 operators are not larger than the diagonal ones in
size. Then, $V^{\ell}_{L}$ and $V^{\ell}_{R}$ can be obtained by diagonalizing
the matrices $U^{\dag}_{\omega}m_{\ell}m^{\dag}_{\ell}U_{\omega}$ and
$m^{\dag}_{\ell}m_{\ell}$, respectively, in Eq.~(\ref{CHcorrect}).
Especially, the mixing matrix $V^{\ell}_{L}$ takes part in PMNS mixing matrix. The matrix $V^{\ell}_{L}$ can be, in general,
parametrized in terms of three mixing angles and six {\it CP}
violating phases :
 \begin{eqnarray}
 V^{\ell}_{L}={\left(\begin{array}{ccc}
 c_{2}c_{3} &  c_{2}s_{3}e^{i\phi^{\ell}_{3}} &  s_{2}e^{i\phi^{\ell}_{2}} \\
 -c_{1}s_{3}e^{-i\phi^{\ell}_{3}}-s_{1}s_{2}c_{3}e^{i(\phi^{\ell}_{1}-\phi^{\ell}_{2})} &  c_{1}c_{3}-s_{1}s_{2}s_{3}e^{i(\phi^{\ell}_{1}-\phi^{\ell}_{2}+\phi^{\ell}_{3})} &  s_{1}c_{2}e^{i\phi^{\ell}_{1}} \\
 s_{1}s_{3}e^{-i(\phi^{\ell}_{1}+\phi^{\ell}_{3})}-c_{1}s_{2}c_{3}e^{-i\phi^{\ell}_{2}} &  -s_{1}c_{3}e^{-i\phi^{\ell}_{1}}-c_{1}s_{2}s_{3}e^{i(\phi^{\ell}_{3}-\phi^{\ell}_{2})} &  c_{1}c_{2}
 \end{array}\right)}P_{\ell}~,
 \label{Vl}
 \end{eqnarray}
where $s_{i}\equiv \sin\theta_{i}$, $c_{i}\equiv \cos\theta_{i}$
and a diagonal phase matrix $P_{\ell}={\rm
Diag.}(e^{i\xi'_{1}},e^{i\xi'_{2}},e^{i\xi'_{3}})$ which can be
rotated away by redefinition of left-charged fermion fields.
In the charged fermion (quarks and charged leptons) sector, there
is a qualitative feature which distinguishes the neutrino sector
from the charged fermion one. The mass spectrum of the charged
leptons exhibits a similar hierarchical pattern as that of the
down-type quarks, unlike that of the up-type quarks which shows a
much stronger hierarchical pattern. For example, in terms of the
Cabbibo angle $\lambda \equiv \sin\theta_{\rm C} \approx
|V_{us}|$, the fermion masses scale as ~$(m_{e},m_{\mu}) \approx
(\lambda^{5},\lambda^{2})~ m_{\tau}$,~$(m_{d},m_{s}) \approx
(\lambda^{4},\lambda^{2})~ m_{b}$ and $(m_{u},m_{c}) \approx
(\lambda^{8},\lambda^{4})~ m_{t}$.  This may lead to two
implications: (i) the Cabibbo-Kobayashi-Maskawa (CKM) matrix
\cite{CKM} is mainly governed by the down-type quark mixing
matrix, and (ii) the charged lepton mixing matrix is similar to
that of the down-type quark one. Therefore, we shall assume that
(i) $V_{\rm CKM}=V^{d\dag}_{L}$ and $V^{u}_{L}=\mathbf{I}$, where
$V^{d}_{L}~(V^{u}_{L})$ is associated with the diagonalization of
the down-type (up-type) quark mass matrix and $\mathbf{I}$ is a
$3\times3$ unit matrix, and (ii) the charged lepton mixing matrix
$V^{\ell}_{L}$ has the similar structure as the CKM matrix. Now, putting a reasonable assumption
 \begin{eqnarray}
  1\gg \frac{|m^{\ell}_{22}|}{|m^{\ell}_{33}|}\sim \frac{|m^{\ell}_{23}|}{|m^{\ell}_{33}|}\gg\frac{|m^{\ell}_{13}|}{|m^{\ell}_{33}|}\sim \frac{|m^{\ell}_{12}|}{|m^{\ell}_{33}|}\gg \frac{|m^{\ell}_{11}|}{|m^{\ell}_{33}|}\sim \frac{|m^{\ell}_{32}|}{|m^{\ell}_{33}|}\gg \frac{|m^{\ell}_{21}|}{|m^{\ell}_{33}|}\sim \frac{|m^{\ell}_{31}|}{|m^{\ell}_{33}|}
 \label{hierarchych}
 \end{eqnarray}
 into Eq.~(\ref{CHcorrect}),
then the measured mass hierarchy of charged lepton is expressed as $m_{\mu}/m_{\tau}\approx|m^{\ell}_{22}/m^{\ell}_{33}|\approx\lambda^{2},  m_{e}/m_{\tau}\approx|m^{\ell}_{11}/m^{\ell}_{33}|\approx0.6\lambda^{5}$ and $m_{e}/m_{\mu}\approx|m^{\ell}_{11}/m^{\ell}_{22}|\approx0.5\lambda^{3}$.
And the mixing angles and phases in Eq.~(\ref{Vl}) can be roughly expressed as
 \begin{eqnarray}
  \theta_{1}&\simeq&\frac{|m^{\ell}_{23}|}{|m^{\ell}_{33}|}~,\qquad\qquad \quad \phi_{1}\simeq \frac{1}{2}\arg(m^{\ell}_{22}m^{\ell\ast}_{32}+m^{\ell}_{23}m^{\ell\ast}_{33})~,  \nonumber \\ \theta_{2}&\simeq&\frac{|m^{\ell}_{13}|}{|m^{\ell}_{33}|}~,\qquad\qquad  \quad\phi_{2}\simeq \frac{1}{2}\arg(m^{\ell}_{11}m^{\ell\ast}_{31}+m^{\ell}_{13}m^{\ell\ast}_{33})~,   \nonumber \\ \theta_{3}&\simeq&\frac{|m^{\ell}_{12}|}{|m^{\ell}_{22}|}~, \qquad\qquad \quad \phi_{3}\simeq\frac{1}{2}\arg(m^{\ell}_{11}m^{\ell\ast}_{21}+m^{\ell}_{12}m^{\ell\ast}_{22}) \ .
 \end{eqnarray}
Now letting $|(V^{\ell}_{L})_{12}|\equiv|c_{2}s_{3}|\approx\lambda$ in similar to the case of quark sector~\cite{Ahn:2011yj}, we obtain $|(V^{\ell}_{L})_{12}|\approx|m^{\ell}_{12}|/|m^{\ell}_{22}|\approx\lambda^{-2}|m^{\ell}_{12}|/|m^{\ell}_{33}|$, leading to $|m^{\ell}_{12}|/|m^{\ell}_{33}|\approx\lambda^{3}$. Consequently, $\theta_{1}\simeq|m^{\ell}_{23}|/|m^{\ell}_{33}|\simeq\lambda^{2}$ and $\theta_{2}\simeq|m^{\ell}_{13}|/|m^{\ell}_{33}|\simeq\lambda^{3}$ are obtained from Eq.~(\ref{hierarchych}). Then, the mixing matrix $V^{\ell}_{L}$ in Eq.~(\ref{Vl}) can be written as
 \begin{eqnarray}
 V^{\ell}_{L}= {\left(\begin{array}{ccc}
 1-\frac{\lambda^{2}}{2} & \lambda e^{i\phi^{\ell}_{3}}  & A\lambda^{3}e^{i\phi^{\ell}_{2}}  \\
 -\lambda e^{-i\phi^{\ell}_{3}} &  1-\frac{\lambda^{2}}{2} & B\lambda^{2}e^{i\phi^{\ell}_{1}} \\
 -A\lambda^{3}e^{-i\phi^{\ell}_{2}}+B\lambda^{3}e^{-i(\phi^{\ell}_{1}+\phi^{\ell}_{3})} & -B\lambda^{2}e^{-i\phi^{\ell}_{1}} & 1
 \end{array}\right)}P_{\ell}+{\cal O}(\lambda^{4})~,
 \label{LL}
 \end{eqnarray}
where the coefficients $A$ and $B$ are real and positive coefficients, but less than one in magnitude.

The Yukawa interactions in Eq.~(\ref{lagrangian}) and the charged gauge interactions in a weak eigenstate basis can be written as
 \begin{eqnarray}
 -{\cal L} &=& \frac{1}{2}\overline{N^{c}_{R}}M_{R}N_{R}+\overline{\ell_{L}}m_{\ell}\ell_{R}
 +\overline{\nu_{L}}Y_{\nu}N_{R}\eta+\frac{g}{\sqrt{2}}W^{-}_{\mu}\overline{\ell_{L}}\gamma^{\mu}\nu_{L}+h.c~.
 \label{lagrangianA}
 \end{eqnarray}
When dealing with lepton flavor violation and DM it is convenient to work in the basis where heavy Majorana neutriino and charged lepton mass matrices are diagonal. In order to go into the physical basis (mass basis) of the right-handed neutrino, performing basis rotations
 \begin{eqnarray}
 L_{L}\rightarrow V^{\ell\dag}_{L}U^{\dag}_{\omega}L_{L}~,~~~~~\ell_{R}\rightarrow V^{\dag}_{R}\ell_{R}~,~~~~N_{R}\rightarrow U^{\dag}_{R}N_{R}
 \label{basis}
 \end{eqnarray}
where $L_{L}=(\nu_{L},\ell_{L})^{T}$, so that the Yukawa coupling matrix $Y_{\nu}$ gets modified to
 \begin{eqnarray}
 Y_{\nu}\rightarrow \tilde{Y}_{\nu}=V^{\ell\dag}_{L}U^{\dag}_{\omega}Y_{\nu}U_{R}~,
 \label{YnuT}
 \end{eqnarray}
and both the right-handed Majorana mass matrix $M_{R}$ and the charged lepton matrix $m_{\ell}$ become diagonal by the unitary matrix $U_{R}$ and $U_{\omega}V^{\ell}_{L}$, respectively;
 \begin{eqnarray}
  \hat{m}_{\ell}=V^{\ell\dag}_{L}U^{\dag}_{\omega}m_{\ell}V^{\ell}_{R}={\rm Diag.}(m_{e},m_{\mu},m_{\tau})~,~~~~\hat{M}_{R}=U^{T}_{R}M_{R}U_{R}=M{\rm Diag.}(a,1,b)
  \label{heavy}
 \end{eqnarray}
where $a=\sqrt{1+\kappa^{2}+2\kappa\cos\xi}$ and $b=\sqrt{1+\kappa^{2}-2\kappa\cos\xi}$ with real and positive mass eigenvalues, $M_{1}=Ma, M_{2}=M$ and $M_{3}=Mb$, and the diagonalizing matrix $U_{R}$ is
\begin{eqnarray}
  U_{R} = \frac{1}{\sqrt{2}} {\left(\begin{array}{ccc}
  0  &  \sqrt{2}  &  0 \\
  1 &  0  &  -1 \\
  1 &  0  &  1
  \end{array}\right)}{\left(\begin{array}{ccc}
  e^{i\frac{\psi_1}{2}}  &  0  &  0 \\
  0  &  1  &  0 \\
  0  &  0  &  e^{i\frac{\psi_2}{2}}
  \end{array}\right)}~,
  \label{URN}
\end{eqnarray}
with the phases
\begin{eqnarray}
 \psi_1 = \tan^{-1} \Big( \frac{\kappa\sin\xi}{1+\kappa\cos\xi} \Big)
 ~~~{\rm and}~~~ \psi_2 = \tan^{-1} \Big( \frac{\kappa\sin\xi}{\kappa\cos\xi-1} \Big)~.
\label{alphs_beta}
\end{eqnarray}

\section{Low energy observables}
\begin{figure}[t]
\begin{center}
\includegraphics*[width=0.4\textwidth]{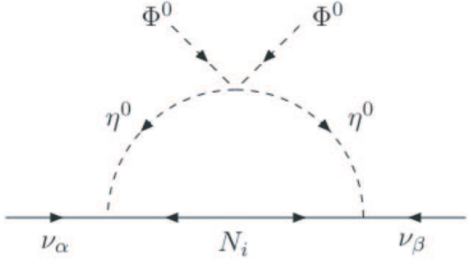}
\caption{\label{Fig1} One-loop generation of light neutrino masses.}
\end{center}
\end{figure}
We now proceed to discuss the low energy neutrino observables.
Due to the $Z_{2}$ symmetry, we can not get the Yukawa Dirac neutrino mass matrix and therefore the usual seesaw does not operate any more, however, similar to~\cite{Ma:2006fn} the light neutrino mass matrix can be generated through one loop diagram in Fig.~\ref{Fig1} due to the quartic scalar interactions. After electroweak symmetry breaking, i.e. $\langle\Phi^{0}\rangle=v\mathbf{I}$, in a basis where charged lepton mass matrix is real diagonal the flavor neutrino masses can be written as
 \begin{eqnarray}
   (m_{\nu})_{\alpha\beta}=\frac{\Delta m^{2}_{\eta}}{16\pi^{2}}\sum_{i}\frac{(\tilde{Y}_{\nu})_{\alpha i}(\tilde{Y}_{\nu})_{\beta i}}{M_{i}}f\left(\frac{M^{2}_{i}}{\bar{m}^{2}_{\eta}}\right)
 \label{lownu1}
 \end{eqnarray}
where
 \begin{eqnarray}
   f(z_{i})=\frac{z_{i}}{1-z_{i}}\left[1+\frac{z_{i}\ln z_{i}}{1-z_{i}}\right]~,\qquad \Delta m^{2}_{\eta}\equiv|m^{2}_{R}-m^{2}_{I}|=6\lambda^{\Phi\eta}_{3}v^{2}
 \label{lambda}
 \end{eqnarray}
with $z_{i}=M^{2}_{i}/\bar{m}^{2}_{\eta}$, if $m_{R}(m_{I})$ is the mass of $\eta^{0}_{R}(\eta^{0}_{I})$ and $m^{2}_{R(I)}=\bar{m}^{2}_{\eta}\pm\Delta m^{2}_{\eta}/2$ where the subscripts $R(I)$ indicate real (imaginary) component\footnote{See Appendix.}, respectively. With $\tilde{M}_{R}={\rm Diag}(M_{r1},M_{r2},M_{r3})$ and $M_{ri}\equiv M_{i}f^{-1}(z_{i})$, the above formula Eq.~(\ref{lownu1}) can be expressed as
 \begin{eqnarray}
  m_{\nu} &=& \frac{\Delta m^{2}_{\eta}}{16\pi^{2}}\tilde{Y}_{\nu}\tilde{M}^{-1}_{R}\tilde{Y}^{T}_{\nu}\nonumber\\
  &=& V^{\ell\dag}_{L}U^{\dag}_{\omega}U_{\nu}~{\rm Diag.}(m_{1},m_{2},m_{3})~U^{T}_{\nu}U^{\ast}_{\omega}V^{\ell\ast}_{L}=m_{0}V^{\ell\dag}_{L}U^{\dag}_{\omega}{\left(\begin{array}{ccc}
  f(z_{2}) & 0 & 0 \\
  0 & A & G  \\
  0 & G & B
 \end{array}\right)}U^{\ast}_{\omega}V^{\ell\ast}_{L}~,
 \label{radseesaw3}
 \end{eqnarray}
where $m_{i}(i=1,2,3)$ are the light neutrino mass eigenvalues and
 \begin{eqnarray} A&=&f(z_{1})\frac{(1+y_{1}e^{i\rho_{1}})^{2}e^{i\psi_{1}}}{2a}+f(z_{3})\frac{(1-y_{1}e^{i\rho_{1}})^{2}e^{i\psi_{2}}}{2b}~,\qquad \qquad  m_{0}= \frac{\Delta m^{2}_{\eta}|y_{\nu}|^{2}}{16\pi^{2}M}~,\nonumber\\
  B&=&f(z_{1})\frac{(1+y_{2}e^{i\rho_{2}})^{2}e^{i\psi_{1}}}{2a}+f(z_{3})\frac{(1-y_{2}e^{i\rho_{2}})^{2}e^{i\psi_{2}}}{2b} ~,\nonumber\\
  G&=&f(z_{1})\frac{(1+y_{1}e^{i\rho_{1}})(1+y_{2}e^{i\rho_{2}})e^{i\psi_{1}}}{2a}-f(z_{3})\frac{(1-y_{1}e^{i\rho_{1}})(1-y_{2}e^{i\rho_{2}})e^{i\psi_{2}}}{2b}~.
 \label{entries}
 \end{eqnarray}

As can be seen in Eq.~(\ref{radseesaw3}), the leptonic mixing matrix is given as
 \begin{eqnarray}
  U_{\rm PMNS}=V^{\ell\dag}_{L}U^{\dag}_{\omega}U_{\nu}~,
 \label{PMNS}
 \end{eqnarray}
with the additional mixing matrix $U_{\nu}$ which is given by
 \begin{eqnarray}
  U_{\nu}={\left(\begin{array}{ccc}
  1 & 0 & 0 \\
  0 & e^{i\varphi_{1}} & 0  \\
  0 & 0 & e^{i\varphi_{2}}
 \end{array}\right)}{\left(\begin{array}{ccc}
  0 & 1 & 0 \\
  \cos\theta & 0 & \sin\theta  \\
  -\sin\theta & 0 & \cos\theta
 \end{array}\right)}P_{\nu}~,
 \label{Unu}
 \end{eqnarray}
where the phase matrix of Majorana particle $P_{\nu}={\rm Diag.}(e^{i\zeta_{1}},e^{i\zeta_{2}},e^{i\zeta_{3}})$ can be absorbed into the neutrino mass eigenstates fields.
The phase $\varphi_{12}$ and the mixing angle $\theta$ are given by
 \begin{eqnarray}
  \varphi_{12}\equiv\varphi_{1}-\varphi_{2}=\arg\left\{AG^{\ast}+GB^{\ast}\right\}~,\qquad \tan2\theta=\frac{2|AG^{\ast}+GB^{\ast}|}{|B|^{2}-|A|^{2}}~,
 \end{eqnarray}
indicating that, in the limit of $y_{1,2}$ approaching to zero, the angle $\theta$ goes to $\pi/4(-\pi/4)$ and the phase $\varphi_{12}$ goes to $\pi(0)$ for $\frac{f^{2}(z_{1})}{a^{2}}<\frac{f^{2}(z_{3})}{b^{2}}~\left(\frac{f^{2}(z_{1})}{a^{2}}>\frac{f^{2}(z_{3})}{b^{2}}\right)$, whose one-to-one correspondence comes from the constraint of mixing parameters presented in Eq.~(\ref{Mangle}) with experimental data, especially see the parameter $\varepsilon$ in Eq.~(\ref{12parameterI}).

The light neutrino mass eigenvalues are given as
 \begin{eqnarray}
  m^{2}_{1}&=&m^{2}_{0}\left(|G|^{2}+|A|^{2}\cos^{2}\theta+|B|^{2}\sin^{2}\theta-|AG^{\ast}+GB^{\ast}|\sin2\theta\right)~,\nonumber\\
  m^{2}_{2}&=&m^{2}_{0}f^{2}(z_{2})~,\nonumber\\
  m^{2}_{3}&=&m^{2}_{0}\left(|G|^{2}+|A|^{2}\sin^{2}\theta+|B|^{2}\cos^{2}\theta+|AG^{\ast}+GB^{\ast}|\sin2\theta\right)~.
 \label{masseigen}
 \end{eqnarray}
Because of the observed hierarchy $|\Delta m^{2}_{\rm Atm}|\gg\Delta m^{2}_{\rm Sol}>0$, and the requirement of Mikheyev-Smirnov-Wolfenstein resonance for solar neutrinos, there are two possible neutrino mass spectrum: (i) $m_{1}<m_{2}<m_{3}$ (normal mass spectrum) which corresponds to $\theta=\pi/4+\delta$ and $\varphi_{12}=\pi+\delta'$, and (ii) $m_{3}<m_{1}<m_{2}$ (inverted mass spectrum) which corresponds to $\theta=-\pi/4+\delta$ and $\varphi_{12}=\delta'$, where $|\delta|,|\delta'|\ll1$. The solar and atmospheric mass-squared differences are given by
\begin{eqnarray}
 \Delta m^{2}_{\rm Sol}\equiv m^{2}_{2}-m^{2}_{1}&=& m^{2}_{0}\left\{f^{2}(z_{2})-|G|^{2}-|A|^{2}\cos^{2}\theta-|B|^{2}\sin^{2}\theta+|AG^{\ast}+GB^{\ast}|\sin2\theta\right\}~,\nonumber\\
 \Delta m^{2}_{\rm Atm}\equiv m^{2}_{3}-m^{2}_{1}&=& 2m^{2}_{0}\frac{|AG^{\ast}+GB^{\ast}|}{\sin2\theta}~,
 \label{deltam2}
\end{eqnarray}
which are constrained by the neutrino oscillation experimental results.
Note here that the parameter $M_{ri}$ can be simplified in the following limit cases as
 \begin{eqnarray}
  M_{ri}\simeq\left\{
  \begin{array}{ll}
    M_{i}\left[\ln z_{i}-1\right]^{-1}, & \hbox{for $z_{i}\gg1$} \\
    2M_{i}, & \hbox{for $z_{i}\rightarrow1$} \\
    \bar{m}^{2}_{\eta}M^{-1}_{i}, & \hbox{for $z_{i}\ll1$~.}
  \end{array}
\right.
 \end{eqnarray}
In this work we will focus on the case the lightest $Z_{2}$-odd neutral particle of $N_{i}$ which is stable and can be a candidate of DM.
As will be seen in section-\ref{DF}, DM constraint (Br$(\mu\to e\gamma)+$WMAP analysis) gives $m_{\eta^{\pm}}\approx M_{\rm lightest}$ where $M_{\rm lightest}$ is the lightest of the heavy Majorana neutrinos. And, due to $|\lambda^{\Phi\eta}_{2}|\lesssim4\pi$, one can obtain $M_{\rm lightest}\approx\bar{m}_{\eta}$ (see Appendix). From Eq.~(\ref{heavy}), depending on the hierarchy of the heavy Majorana neutrino masses $M_{1},M_{2}$ and $M_{3}$, the relative size of the parameter $\kappa$ consistent with the possible mass ordering of light neutrinos and hierarchy of $\Delta m^{2}_{32}$ and $\Delta m^{2}_{21}$ can be classified as follows :

\vskip 0.4cm
{\bf (i)} $M_{3}<M_{2}<M_{1}$ for $\xi=0$:
\begin{itemize}
  \item $M_{3}\simeq\bar{m}_{\eta}\ll M_{2}<M_{1}$ (or $a>1\gg b$) :
  this case corresponds to the normal hierarchical mass spectrum with $b\to0$ {\it i.e.} $\kappa\simeq1$ and $a\simeq2$. Using $f(z_{1})\simeq2\ln(2/b)-1>f(z_{2})\simeq-2\ln b-1\gg f(z_{3})\simeq 1/2$, the condition $\Delta m^{2}_{21}>0$ is obtained for $\theta=+\pi/4$, indicating the normal mass ordering of light neutrinos with $\Delta m^{2}_{32}\approx m^{2}_{0}\left(f^{2}(z_{3})/b^{2}-f^{2}(z_{1})/4\right)>0$. The ratio of the mass squared differences defined by $R\equiv\Delta m^{2}_{21}/|\Delta m^{2}_{32}|$ is given by
 \begin{eqnarray}
  R\approx b^{2}(4f^{2}(z_{2})-f^{2}(z_{1}))~,
 \label{R1}
 \end{eqnarray}
where the equality roughly is given under $y_{1,2},b\ll1$. From Eq.~(\ref{R1}) and the best-fit values of the solar and atmospheric mass squared differences ($R\simeq3\times10^{-2}$), one can determine the magnitude of the parameter $b$, roughly
 \begin{eqnarray}
  b\approx0.01~.
 \label{R11}
 \end{eqnarray}
  \item $M_{3}\simeq\bar{m}_{\eta}<M_{2}<M_{1}$ (or $b<1<a$) : this case gives $a\simeq1.4$, $b\simeq0.6$ with $\kappa\simeq0.4$ and $a\simeq2.8$, $b\simeq0.8$ with $\kappa\simeq1.8$ in numerical calculations, which corresponds to $f(z_{3})/b\lesssim f(z_{1})/a\lesssim f(z_{2})$ indicating a degenerate inverted orderings of light neutrinos.
\end{itemize}

\vskip 0.4cm
{\bf (ii)} $M_{1}\simeq\bar{m}_{\eta}<M_{2}<M_{3}$ for $\xi=\pi$ : this case $a\lesssim1<b$ corresponds to a degenerate inverted ordering of light neutrinos giving $a\simeq1$ and $b\simeq3$ with $\kappa\simeq2$ in numerical calculations.
Note that a case $M_{1}\simeq\bar{m}_{\eta}\ll M_{2}<M_{3}$ is not allowed, due to $\varepsilon\to2$ in Eq.~(\ref{Mangle}), actually because it could not satisfy the experimental data of light neutrino of mixing angles.

\vskip 0.4cm
{\bf (iii)} $M_{2}\simeq\bar{m}_{\eta}< M_{1}< M_{3}~(1<b<a)$ for $\xi=0$ :
 this corresponds to $f(z_{1})/a\lesssim f(z_{2})<f(z_{3})/b$, giving $a\simeq4.7$ and $b\simeq2.7$ with $\kappa\simeq3.7$ in numerical calculations. It gives a degenerate normal mass ordering of light neutrinos.
Note that a case $M_{2}\simeq\bar{m}_{\eta}\ll M_{1}\simeq M_{3}$ for $\xi\simeq\pi/2~{\rm or}~\xi\simeq3\pi/2$ is not allowed because it could not satisfy the ratio $R\simeq3\times10^{-3}$.

Note here that in our scenario the inverted hierarchical light neutrino mass spectrum is not allowed because the condition $\Delta m^{2}_{21}>0$ is not satisfied due to the mass ordering of heavy Majorana neutrinos Eq.~(\ref{heavy}) corresponding to light neutrino mass ordering.

From Eqs.~(\ref{CHcorrect}), (\ref{LL}) and (\ref{Unu}), the PMNS matrix in Eq.~(\ref{PMNS})
can be expressed as
 \begin{eqnarray}
 \small
  U_{\rm PMNS}=
   {\left(\begin{array}{ccc}
  V^{\ell}_{L11}V_{11}-V^{\ell}_{L12}V_{21}+V^{\ell\ast}_{L31}V_{31}
   & \frac{V^{\ell}_{L11}-V^{\ell}_{L12}+V^{\ell\ast}_{L31}}{\sqrt{3}}
   & V^{\ell}_{L11}V_{13}-V^{\ell}_{L12}V_{23}+V^{\ell\ast}_{L31}V_{33} \\
  V^{\ell}_{L22}V_{21}-V^{\ell}_{L23}V_{31}+V^{\ell\ast}_{L12}V_{11}
   & \frac{V^{\ell}_{L22}-V^{\ell}_{L23}+V^{\ell\ast}_{L12}}{\sqrt{3}}
   &  V^{\ell}_{L22}V_{23}-V^{\ell}_{L23}V_{33}+V^{\ell\ast}_{L12}V_{13} \\
  V^{\ell}_{L31}V_{33}+V^{\ell\ast}_{L13}V_{11}+V^{\ell\ast}_{L23}V_{21}
   & \frac{V^{\ell}_{L33}+V^{\ell\ast}_{L13}+V^{\ell\ast}_{L23}}{\sqrt{3}}
   & V^{\ell}_{L33}V_{33}+V^{\ell\ast}_{L13}V_{13}+V^{\ell\ast}_{L23}V_{23}
 \end{array}\right)}P_{\nu},
  \label{PMNS2}
 \end{eqnarray}
where $V^{\ell}_{Lij}$ is the $(ij)$-element of the mixing matrix $V^{\ell}_L$, and $V_{ij}$ is the
$(ij)$-element of $U^{\dag}_{\omega}U_{\nu}$ given by
 \begin{eqnarray}
 V = U^{\dag}_{\omega}U_{\nu}={\left(\begin{array}{ccc}
 \frac{ce^{i\varphi_{1}}-se^{i\varphi_{2}}}{\sqrt{3}} & \frac{1}{\sqrt{3}}
  &  \frac{ce^{i\varphi_{2}}+se^{i\varphi_{1}}}{\sqrt{3}} \\
 -\frac{ce^{i(\varphi_{1}+\frac{\pi}{3})}-se^{i(\varphi_{2}-\frac{\pi}{3})}}{\sqrt{3}}
  &  \frac{1}{\sqrt{3}}
  &  -\frac{se^{i(\varphi_{1}+\frac{\pi}{3})}+ce^{i(\varphi_{2}-\frac{\pi}{3})}}{\sqrt{3}} \\
 -\frac{ce^{i(\varphi_{1}-\frac{\pi}{3})}-se^{i(\varphi_{2}+\frac{\pi}{3})}}{\sqrt{3}}
  &  \frac{1}{\sqrt{3}}
  &  -\frac{se^{i(\varphi_{1}-\frac{\pi}{3})}+ce^{i(\varphi_{2}+\frac{\pi}{3})}}{\sqrt{3}}
 \end{array}\right)}P_{\nu}~,
 \label{PMNS1}
 \end{eqnarray}
where $s\equiv\sin\theta$ and $c\equiv\cos\theta$. By recasting Eq.~(\ref{PMNS2}) with the transformations $e \to e ~e^{i\alpha_{1}}$,
$\mu \to \mu ~e^{i\beta_{1}}$, $\tau \to \tau ~e^{i\beta_{2}}$ and
$\nu_{2} \to \nu_{2} ~e^{i(\alpha_{1}-\alpha_{2})}$, we can rewrite the PMNS matrix as
 \begin{eqnarray}
  U_{\rm PMNS}=
 {\left(\begin{array}{ccc}
 |U_{e1}| & |U_{e2}| & U_{e3}e^{-i\alpha_{1}} \\
 U_{\mu1}e^{-i\beta_{1}} & U_{\mu2}e^{i(\alpha_{1}-\alpha_{2}-\beta_{1})} &  |U_{\mu3}| \\
 U_{\tau1}e^{-i\beta_{2}} & U_{\tau2}e^{i(\alpha_{1}-\alpha_{2}-\beta_{2})} & |U_{\tau3}|
 \end{array}\right)}P'_{\nu}~
 \label{PMNS12}
 \end{eqnarray}
which corresponds to the standard parametrization as in PDG~\cite{pdg}.
From the above equation, the neutrino mixing parameters can be displayed as
 \begin{eqnarray}
  \sin^{2}\theta_{12}&=&\frac{|U_{e2}|^{2}}{1-|U_{e3}|^{2}}~,~~~
   \sin^{2}\theta_{23}=\frac{|U_{\mu3}|^{2}}{1-|U_{e3}|^{2}}~,\nonumber\\
  \sin^{2}\theta_{13}&=&|U_{e3}|^{2}~,~~~~~~~~~~
   \delta_{CP}=\alpha_{1}-\alpha_{3}~,
 \label{mixing1}
 \end{eqnarray}
where $\alpha_{1} = \arg(U_{e1})$, $\alpha_{2} = \arg(U_{e2})$, $\alpha_{3} = \arg(U_{e3})$,
$\beta_{1} = \arg(U_{\mu3})$ and $\beta_{2} = \arg(U_{\tau3})$. From the form of $U_{\rm PMNS}$ given in Eq.~(\ref{PMNS1}), the solar, atmospheric and reactor neutrino mixing angles $\theta_{12},\theta_{23}$ and $\theta_{13}$
can be approximated, up to order $\lambda^{3}$, as
 \begin{eqnarray}
  \sin^{2}\theta_{12} &=& \frac{1-2\lambda\cos\phi^{\ell}_{3}
   +\lambda^{3}(\cos\phi^{\ell}_{3}-2A\cos\phi^{\ell}_{2}
   +2B\cos\tilde{\phi}^{\ell}_{13})}
   {3-\varepsilon-\Xi\lambda+\lambda^{2}\Psi+\Omega\lambda^{3}}~,\nonumber\\
  \sin^{2}\theta_{23}&=&\frac{1-\sin2\theta\cos(\varphi_{12}-\pi/3)
   -\Xi\lambda+\lambda^{2}\Upsilon+\lambda^{3}\Theta}{3-\varepsilon
   -\Xi\lambda+\lambda^{2}\Psi+\lambda^{3}\Omega} ~,\nonumber\\
  \sin\theta_{13}
  &=&\frac{1}{\sqrt{3}}\sqrt{\varepsilon
   +\Xi\lambda-\lambda^{2}\Psi-\lambda^{3}\Omega}~,
  \label{Mangle}
 \end{eqnarray}
where $\tilde{\phi}^{\ell}_{ij}\equiv\phi^{\ell}_{i}+\phi^{\ell}_{j}$ and the parameters $\Psi$
and $\Omega$ are defined as
 \begin{eqnarray}
  \varepsilon&=&1+\sin2\theta\cos\varphi_{12}~,\nonumber\\
  \Psi&=&\sqrt{3}\sin2\theta\cos(\varphi_{12}-\pi/6)~,\nonumber\\
  \Omega&=&\Theta +A \Big[ \sqrt{3}\cos2\theta\sin\phi^{\ell}_{2}
   -\cos\phi^{\ell}_{2}(1+2\sin2\theta\cos(\varphi_{12}-\pi/3)) \Big] ~,\nonumber\\
  \Xi &=&\cos\phi^{\ell}_{3}+\sqrt{3}\sin\phi^{\ell}_{3}\cos2\theta
   +2\sin2\theta\cos\phi^{\ell}_{3} \cos\big(\varphi_{12}+\frac{\pi}{3}\big)~,\nonumber\\
  \Upsilon&=&\Psi+B \Big[ \cos\phi^{\ell}_{1}(1-2\cos\varphi_{12}\sin2\theta)
   +\sqrt{3}\sin\phi^{\ell}_{1}\cos2\theta \Big] ~,
 \label{12parameterI}
 \end{eqnarray}
with
 \begin{eqnarray}
 \Theta= \frac{\Xi}{2}
  +B \Big[ \cos\tilde{\phi}^{\ell}_{13}(1+2\sin2\theta\cos(\varphi_{12}-\pi/3))
   -\sqrt{3}\cos2\theta\sin\tilde{\phi}^{\ell}_{13} \Big]~.
 \end{eqnarray}
In Eq.~(\ref{Mangle}), if we turn off the higher dimensional operators in the Lagrangian, that is, if $\theta \to \pm \pi/4$ and $\varphi_{12} \to \pi(0)$, and $\lambda\to0$, the TBM angles $\sin^{2}\theta_{12}=1/3,\sin^{2}\theta_{23}=1/2$ and $\sin\theta_{13}=0$ are restored, as expected. In the limit of $\theta \to \pi/4$ and $\varphi_{12} \to \pi$ (normal hierarchy of the neutrino masses), or $\theta \to -\pi/4$ and $\varphi_{12} \to 0$ (inverted hierarchy of the neutrino masses), the parameters behave as $\varepsilon\to0$, $\Xi\to0$, $\Psi\to-3/2$, $\Omega\to0$, $\Upsilon\to-3/2-3B\cos\phi^{\ell}_{1}$ and $\Theta\to0$. Then, the neutrino mixing angles can be simplified as
\begin{eqnarray}
 \sin^{2}\theta_{12}\approx\frac{1}{3}-\frac{2\lambda\cos\phi^{\ell}_{3}}{3},~ \sin^{2}\theta_{23}\approx\frac{1}{2}+\left(\cos\phi^{\ell}_{1}-\frac{1}{4}\right)\lambda^{2},~\sin\theta_{13} \simeq \sqrt{\frac{\lambda^{2}}{2}
   +\frac{\varepsilon+\Xi\lambda}{3}}~,
 \label{simple}
\end{eqnarray}
where $A,B=1$ are used. Note here that in the limit of $\varepsilon\to0$ and $\Xi\to0$ the mixing parameter $\sin\theta_{13}$ goes to the value $\lambda/\sqrt{2}$. Leptonic CP violation at low energies can be detected through the neutrino oscillations which are sensitive to the Dirac CP-phase, but insensitive to the Majorana CP-phases in
$U_{\rm PMNS}$~\cite{Branco:2002xf}:
the Jarlskog invariant $J_{CP}\equiv{\rm Im}[U_{e1}U_{\mu2}U^{\ast}_{e2}U^{\ast}_{\mu1}]$, where
$U_{\alpha j}$ is an element of the PMNS matrix in Eq.~(\ref{PMNS2}), with $\alpha=e,\mu,\tau$
corresponding to the lepton flavors and $j=1,2,3$ corresponding to the light neutrino mass eigenstates.
To see how both CP phases $\varphi_{12}$ (coming from the neutrino sector) and $\phi^{\ell}_{1,2,3}$
(coming from the charged lepton sector) are correlated with low energy CP violation measurable through
the neutrino oscillations, let us consider the leptonic CP violation parameter $J_{CP}$ :
 \begin{eqnarray}
  J_{CP}&=&\frac{\cos2\theta}{6\sqrt{3}}
   -\frac{\lambda\sqrt{3}}{9}\sin2\theta\sin\phi^{\ell}_{3}\cos(\varphi_{12}-\pi/6) \nonumber\\
  &-&\frac{\lambda^{2}}{3\sqrt{3}}\Big( \cos2\theta
   -B\sin2\theta\sin\phi^{\ell}_{1}\sin\varphi_{12} \Big)\nonumber\\
  &+&\frac{\lambda^{3}\sqrt{3}}{18}\Big[\sin2\theta\cos(\varphi_{12}-\pi/6)\sin\phi^{\ell}_{3}+2A\sin2\theta\sin^{\ell}_{2}\cos(\varphi_{12}+\pi/6)\nonumber\\
  &+&B\Big\{2\cos2\theta\cos\phi^{\ell}_{12}+\sin2\theta\Big(\sqrt{3}\cos\varphi_{12}\sin\phi^{\ell}_{13}\nonumber\\
  &+&\sin\varphi_{12}(\cos\phi^{\ell}_{3}\sin\phi^{\ell}_{1}+3\sin\phi^{\ell}_{3}\cos\phi^{\ell}_{1})\Big)\Big\}\Big]+{\cal O}(\lambda^{4})~.
  \label{JCP}
 \end{eqnarray}
Note that $\sin\phi^{\ell}_{3}$~ appears in the terms of the first
order in $\lambda$, and its value is crucial to satisfy the neutrino data for the solar mixing angle. In particular, for $\theta \to \pm\pi/4$ and $\varphi_{12} \to \pi ~({\rm or}~ 0)$,
we obtain $J_{CP} \simeq \pm\lambda/6$ for $\sin\phi^{\ell}_{3} \simeq \pm 1$.

\begin{figure}[t]
 \begin{minipage}[t]{6.0cm}
  \epsfig{figure=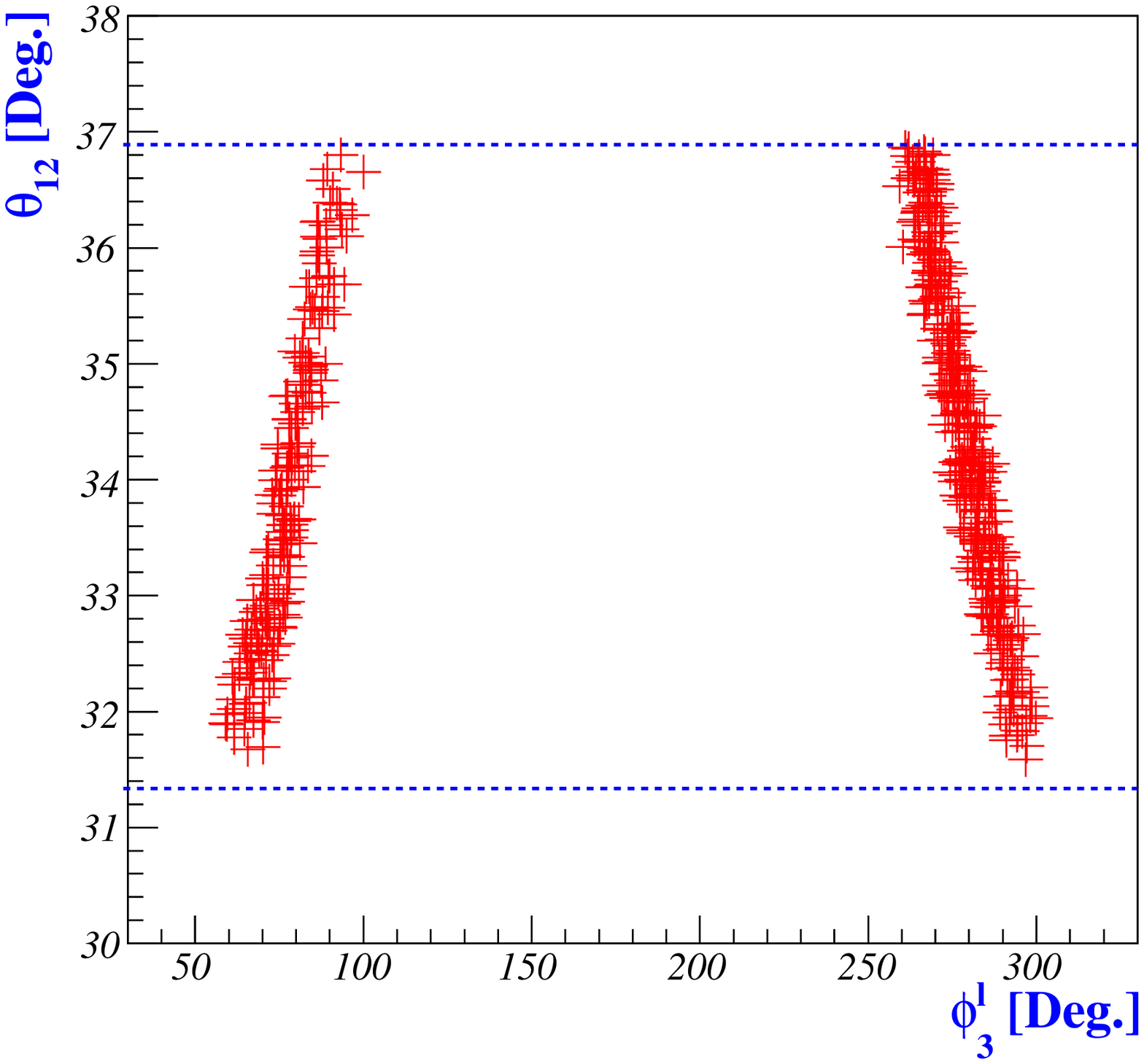,width=6.5cm,angle=0}
 \end{minipage}
 \hspace*{1.0cm}
 \begin{minipage}[t]{6.0cm}
  \epsfig{figure=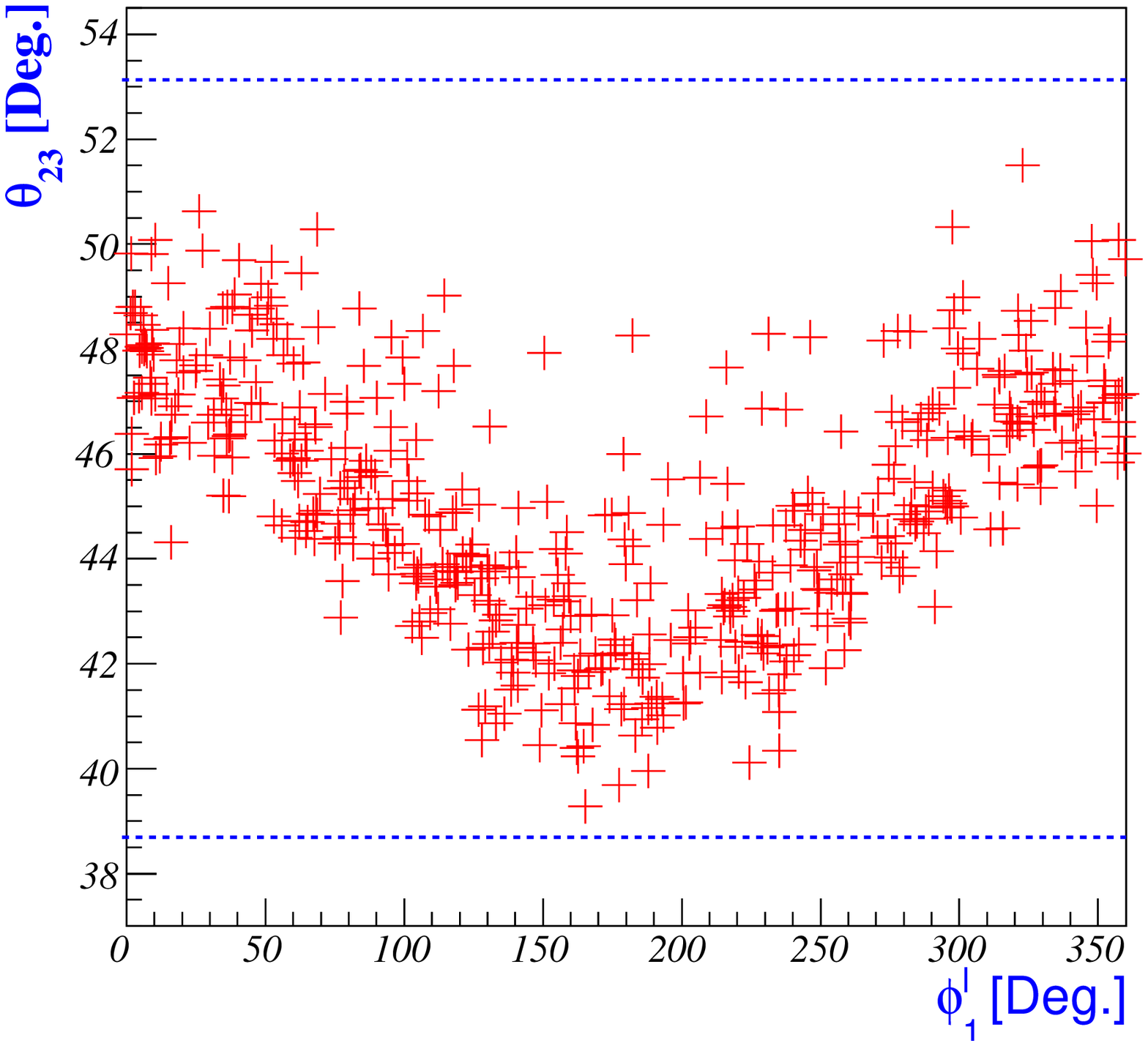,width=6.5cm,angle=0}
 \end{minipage}
 \caption{\label{FigA1}
  Left-plot displays the allowed regions of the solar mixing angle $\theta_{12}$
  versus the CP phase $\phi^{\ell}_{3}$, and right-plot displays the allowed regions of the atmospheric
  mixing angle $\theta_{23}$ versus the CP phase $\phi^{\ell}_{1}$.
  The horizontal dotted lines in both plots correspond to the upper and
  lower bounds, respectively, satisfying the $3\sigma$ experimental data in Eq.~(\ref{expnu}).}
\end{figure}
\begin{figure}[t]
 \begin{minipage}[t]{6.0cm}
  \epsfig{figure=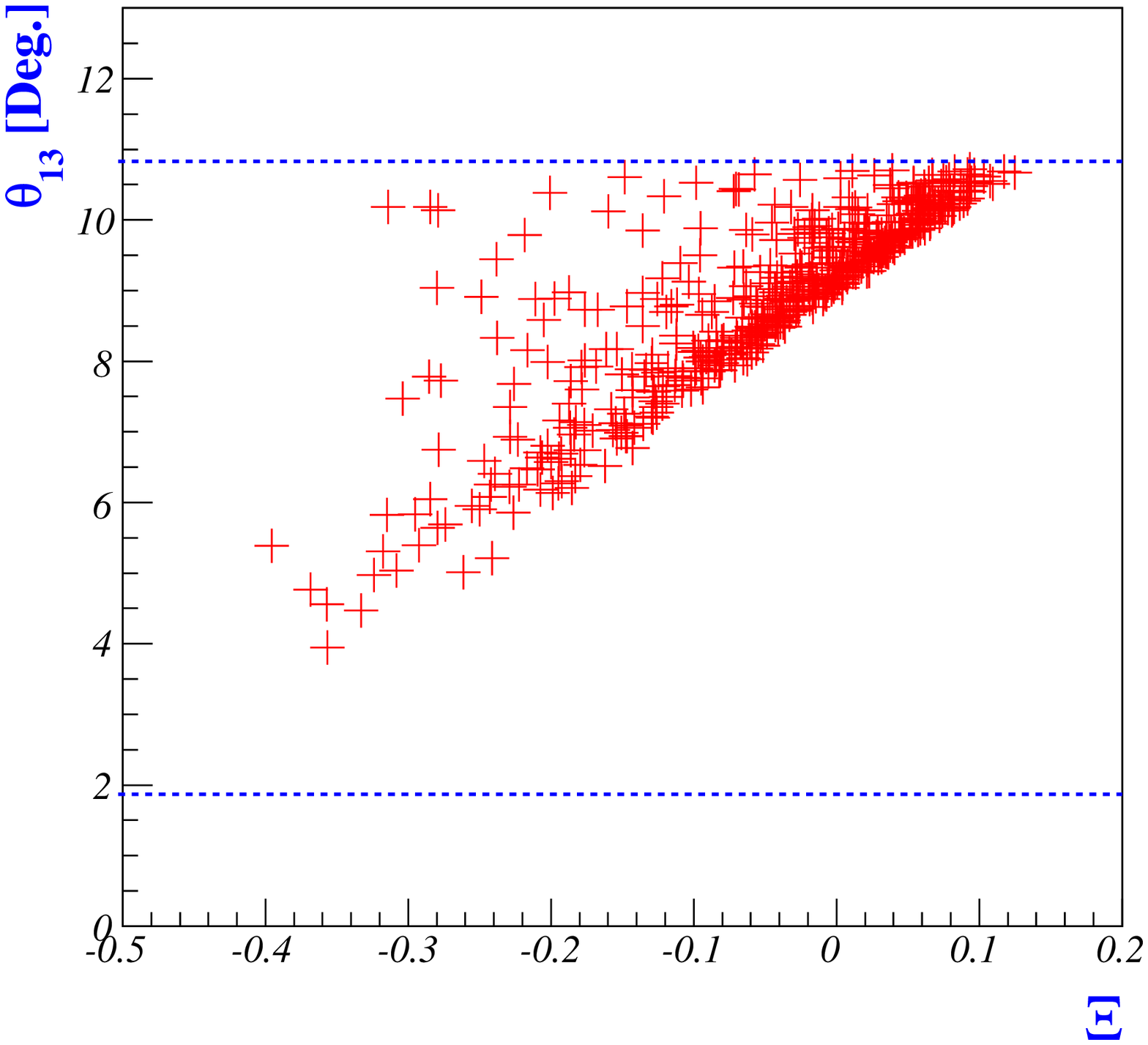,width=6.5cm,angle=0}
 \end{minipage}
 \hspace*{1.0cm}
 \begin{minipage}[t]{6.0cm}
  \epsfig{figure=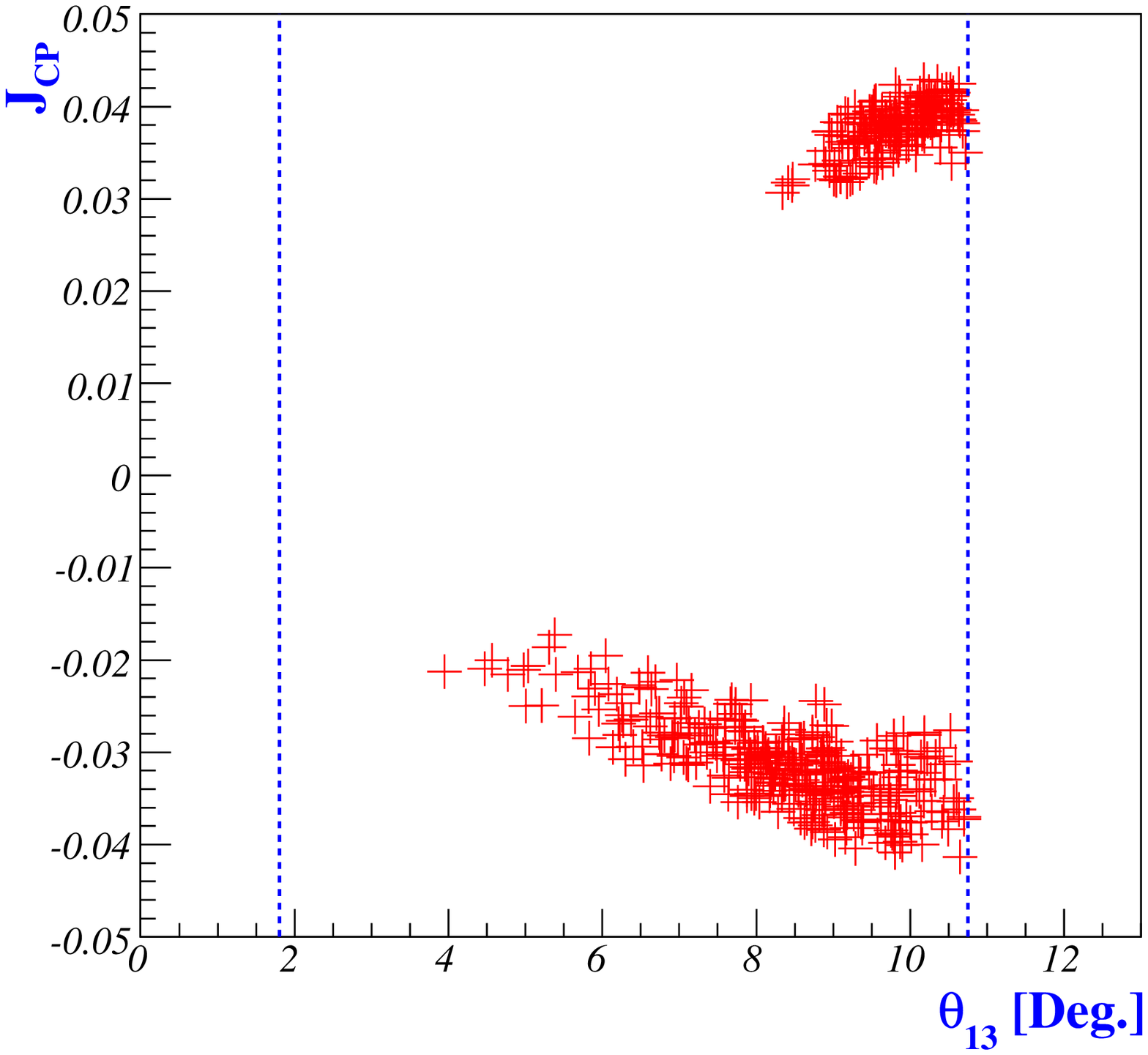,width=6.5cm,angle=0}
 \end{minipage}
 \caption{\label{FigA2}
  Allowed values for the reactor angle $\theta_{13}$  in left-plot and
   $J_{\rm CP}$ in right-plot as a function of $\Xi$ and $\theta_{13}$, respectively.
  The horizontal and vertical dotted lines in the both figures correspond to the upper and lower bounds in $3\sigma$ of experimental data in Eq.~(\ref{expnu}).}
\end{figure}

In our numerical work, we use the five neutrino experimental data of $\Delta m^{2}_{\rm sol}$, $\Delta m^{2}_{\rm atm}$, $\theta_{12},~\theta_{13}$ and $\theta_{23}$ at $3\sigma$ level given in Eq.~(\ref{expnu}) as inputs. On this work, we are going to study the case $M_{3}\ll M_{2}< M_{1}$ giving rise to a normal hierarchical mass ordering of light neutrinos, see Eq.~(\ref{R11}). From Eqs.~(\ref{lambda},\ref{entries},\ref{masseigen}) we obtain
 \begin{eqnarray}
  m_{2}=m_{0}f(z_{2})\simeq\frac{3v^{2}(-2\ln b-1)}{8\pi^{2}}\frac{\lambda^{\Phi\eta}_{3}|y_{\nu}|^{2}}{M}
 \label{m2overall}
 \end{eqnarray}
which should be order of ${\cal O}(0.01)$ eV, indicating $\lambda^{\Phi\eta}_{3}|y_{\nu}|^{2}/M\simeq{\cal O}(10^{-15})~{\rm GeV}^{-1}$.
By using the relation $m_{0}=\Delta m^{2}_{\eta}|y_{\nu}|^{2}/16\pi^{2}M$ with Eqs.~(\ref{lambda},\ref{m2overall}) and the SM Higgs VEV $v=174$ GeV and, for example, by taking the scale $M=10$ TeV and $\lambda^{\Phi\eta}_{3}=10^{-8}$, the values of the relevant parameters satisfying the five neutrino experimental data are taken as
 \begin{eqnarray}
  &&0.98<\kappa<1.02~,\qquad 0.032<|y_{\nu}|<0.045~,\qquad 0.0001<y_{1,2}<0.47~,\nonumber\\
  &&0\leq\phi^{\ell}_{1,2,3}\leq2\pi~,\qquad 0\leq\rho_{1,2}\leq2\pi~.
 \label{parameter}
 \end{eqnarray}
Without loss of generality, we take $A=B=1$ appearing in the charged lepton mixing matrix $V^{\ell}_{L}$, because they do not affect the leptonic mixing mixing parameters significantly.
The left plot in Fig.~\ref{FigA1} shows the allowed region of the solar mixing angle $\theta_{12}$ versus the CP phase $\phi^{\ell}_{3}$. As can be seen the simplified equation of $\sin^{2}\theta_{12}$ in Eq.~(\ref{simple}), the mixing parameter $\theta_{12}$ is dominantly controlled by the phase$\phi^{\ell}_{3}$ allowing $55^{\circ}\leq\phi^{\ell}_{3}\leq100^{\circ}$ and $260^{\circ}\leq\phi^{\ell}_{3}\leq300^{\circ}$.
And the mixing parameter $\theta_{23}$ is dominantly controlled by $\phi^{\ell}_{1}$ as can be seen in Eq.~(\ref{simple}), and the right plot in Fig.~\ref{FigA1} displays the allowed region of the atmospheric mixing angle $\theta_{23}$ versus the CP phase $\phi^{\ell}_{1}$. The left plot of Fig.~\ref{FigA2} shows that the behavior of $\theta_{13}$ as a function of $\Xi$, where there is a lower bound $\theta_{13}\gtrsim3.5^{\circ}$. Because $\varepsilon\geq0$, depending on the sign of $\Xi$, the second term in the squared-root of $\sin\theta_{13}$ in Eq.~(\ref{simple}) can increase or decrease the value of $\theta_{13}$ around the value $\lambda/\sqrt{2}$. Furthermore, because the value of $\Xi$ is bounded by as can be seen in Eq.~(\ref{12parameterI}), we expect that there is a lower bound on the possible value of $\theta_{13}$. The parameter $\Xi$ depends mainly on $y_{1}$ and $y_{2}$, defined in Eq.~(\ref{yukawaNu}), which represent the effects of the dimension-5 operators. Thus, in our scenario the lower bound on the mixing angle $\theta_{13}$ is strongly dependent on the cutoff scale $\Lambda$, the $A_{4}$ symmetry breaking scale $v_{\chi}$ and the relevant couplings $|y^{s,a}_{N}|$, through $y_{1}$ and $y_{2}$. Since neutrino oscillation experiments are sensitive to the Dirac $CP$ phase $\delta_{CP}$, the Jarlskog invariant of the leptonic sector given in Eq.~(\ref{JCP}) would be a signal of $CP$ violation.
The right plot in Fig.~\ref{FigA2} shows that allowed values for $J_{\rm CP}$ as a function of $\theta_{13}$, showing $|J_{CP}|\approx0.01-0.04$ due to the sizable $\theta_{13}$. This can be tested in the future experiments such as the upcoming long baseline neutrino oscillation ones. In the plots Fig.~\ref{FigA1} and Fig.~\ref{FigA2} the horizontal and vertical dotted lines in the both figures correspond to the upper and lower bounds in $3\sigma$ of experimental data in Eq.~(\ref{expnu}).

\section{Dark matter and Lepton Flavor Violation}
\label{DF}
In this section, we study a fermionic DM analysis, that is, right-handed neutrinos. Especially, we are interested in a hierarchical case of heavy Majorana neutrino, $M_{3}\ll M_{2}< M_{1}$ for $\xi=0$, giving a hierarchical normal mass ordering of light neutrinos, as shown in section-III.

The existence of the flavor neutrino mixing implies that the individual lepton charges, $L_{\alpha}, \alpha=e,\mu,\tau$ are not conserved~\cite{Bilenky:1987ty} and processes like $\ell_{\alpha}\rightarrow\ell_{\beta}\gamma$ should take place. Experimental discovery of lepton rare decay processes $\ell_{\alpha}\rightarrow\ell_{\beta}\gamma$ is one of smoking gun signals of physics beyond the SM; thus several experiments have been developed to detect LFV processes. The present experimental upper bounds are given at $90\%$ C.L.~\cite{meg,PDG} as
 \begin{eqnarray}
  {\rm Br}(\mu\rightarrow e\gamma) &\leq&2.4\times10^{-12}~,~~~~~~{\rm Br}(\tau\rightarrow \mu\gamma)\leq4.4\times10^{-8}~,\nonumber\\
  {\rm Br}(\tau\rightarrow e\gamma) &\leq&3.3\times10^{-8}~.
 \label{expLFV}
 \end{eqnarray}

\begin{figure}[htb]
\begin{center}
\includegraphics*[width=0.7\textwidth]{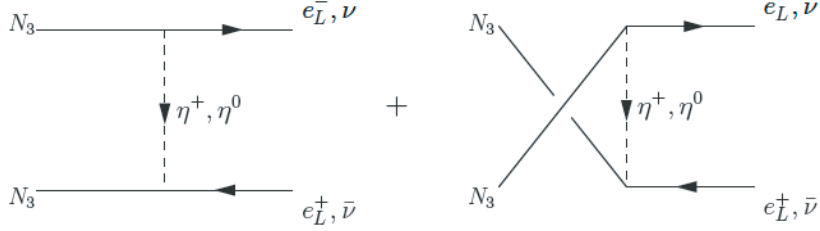}
\end{center}
\caption{\footnotesize
Annihilation diagrams of $n$ for the cross section $\sigma v$, where $e_L$ and $\nu$ run over three families.
}\label{diagrams}
\end{figure}

\begin{figure}[t]
\begin{center}
\includegraphics*[width=0.5\textwidth]{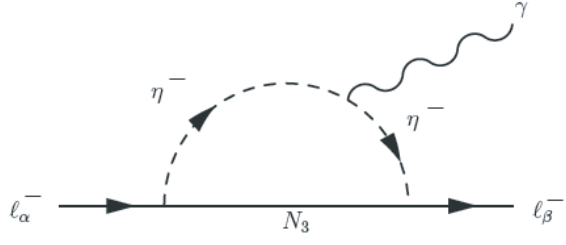}
\caption{One-loop diagrams to the one for neutrino masses contribute to the lepton flavor violating processes like $\ell_{\alpha}\rightarrow\ell_{\beta}\gamma$~$(\alpha,\beta=e,\mu,\tau)$.}
\label{muegamma}
\end{center}
\end{figure}

Assuming the lightest particle of $N_{i}$ to be DM, we consider annihilation of $N_{i}$ through Yukawa interaction Eq.~(\ref{lagrangianA}) in the early Universe. The DM mass $M_{\rm lightest}=M_{3}$ is constrained by the LFV processes, if the size of $|y_{\nu}|$ which is the overall scale of $Y_{\nu}$ is fixed through the DM relic density.
For the given Yukawa interaction of the fermion singlet with SM particles, $\tilde{Y}_{\nu}\bar{\ell}_{L}\eta N_{i}$, its annihilation rate into the latter, as shown in Fig.~\ref{diagrams}, and its relic density $\Omega_{d}$ can be calculated and are related to each other by the thermal dynamics of the Universe within the standard big-bang cosmology~\cite{Kolb}. Then in the WMAP analysis we find the thermally averaged  cross section $\langle \sigma v \rangle$
for the annihilation of two $N_{3}$'s from Fig.~\ref{diagrams} in the limit of the vanishing final state lepton masses:
\begin{eqnarray}
\langle \sigma v\rangle &\simeq&
\frac{|y_\nu|^4 r^2 (1-2 r+2 r^2)}{48 \pi m^2_{\eta^{\pm}}},
\label{bandr}\quad
  r =  M_3^2/(m^2_{\eta^{\pm}}+M_3^2),
  \label{randy}
\end{eqnarray}
where we assume $m_{\eta^0}=m_{\eta^{\pm}}$.
The thermally averaged cross section Eq.(\ref{bandr}) does not contain s-wave contribution as a consequence of massless limit of the final state particles. Assuming this to be the dominant contribution to the DM relic density of the Universe, we need that the pair annihilation cross section must be of the order of
\begin{eqnarray}
 \langle\sigma_{\rm eff}v_{r}\rangle\simeq8.35\times10^{-10}\frac{x_{f}}{\sqrt{g_{\ast}}}\left(\frac{\Omega_{d}h^{2}}{0.11}\right)^{-1}{\rm GeV}^{-2}~,
\label{bandr1}
\end{eqnarray}
which is the thermally averaged annihilation cross section of a pair of fermion particles into SM particles and the relative speed of the fermion pair in their center-of-mass frame, where the decoupling temperature $T_{f}$ of particles, $x_{f}\simeq\ln\frac{0.038M_{\rm Pl}m\langle\sigma_{\rm eff}|v_{r}|\rangle}{\sqrt{g_{\ast}x_{f}}}$, the current Particle Data Group value for the DM density, $\Omega_{d}h^{2}=0.1123\pm0.0035$ \cite{wmap}, in which $\Omega_{d}$ is the cosmological parameter associated with DM, $h$ is the normalized Hubble constant, and $g_{\ast}$ is the number of relativistic degrees of freedom at the DM chemical coupling.
From Eqs.~(\ref{bandr},\ref{bandr1}), one can obtain for $m_{\eta^{\pm}}\equiv x_{\eta}$ GeV
\begin{eqnarray}
 |y_{\nu}|\simeq4.3\times10^{-2}\sqrt{x_{\eta}}\left(\frac{\langle\sigma v\rangle}{3\times10^{-9}{\rm GeV}^{-2}}\right)^{\frac{1}{4}}\left(\frac{m_{\eta^{\pm}}}{x_{\eta}{\rm GeV}}\right)^{\frac{1}{2}}~,
\label{coupling}
\end{eqnarray}
where $r\simeq1/2$ is used. From Eq.~(\ref{coupling}) we see that the coupling depends on the scale of dark matter. For the weak coupling to be ensured, one needs $|y_{\nu}|\lesssim4\pi$
\footnote{$|y_\nu|\le1$ case has been considered in Ref. \cite{other-lfv}. }
which, in turn, indicates the upper bound on the DM mass must be roughly
\begin{eqnarray}
 M_{3}\lesssim83~{\rm TeV}~.
\label{upper}
\end{eqnarray}
Now let us consider LFV, especially $\tau\to\mu\gamma$ for a lower bound of DM and $\mu\to e\gamma$ for a connection to the neutrino parameter $\theta_{13}$.
Fig.~\ref{muegamma} depicts that one-loop diagrams to the one for neutrino masses contribute to the lepton flavor violating processes like $\ell_{\alpha}\rightarrow\ell_{\beta}\gamma$~$(\alpha,\beta=e,\mu,\tau)$, whose branching ratio is estimated as~\cite{Ma:2001mr}
 \begin{eqnarray}
  {\rm Br}(\ell_{\alpha}\rightarrow\ell_{\beta}\gamma)=\frac{3\alpha_{e}}{64\pi(G_{F}m^{2}_{\eta^{\pm}})^{2}}|B_{\alpha\beta}|^{2}{\rm Br}(\ell_{\alpha}\rightarrow\ell_{\beta}\bar{\nu}_{\beta}\nu_{\alpha})
 \label{LFV}
 \end{eqnarray}
where $\alpha_{e}\simeq1/137$ and $G_{F}$ is the Fermi constant, and $B_{\alpha\beta}$ is given by
 \begin{eqnarray}
  B_{\alpha\beta}=\sum^{3}_{i=1}
  \tilde{Y}_{\alpha i}\tilde{Y}^{\ast}_{\beta i}F_{2}(x_{i})~,
 \end{eqnarray}
in which $F_{2}(x_{i})$ is given by
 \begin{eqnarray}
  F_{2}(x_{i})=\frac{1-6x_{i}+3x^{2}_{i}+2x^{3}_{i}-6x^{2}_{i}{\rm ln}x_{i}}{6(1-x_{i})^{4}}
 \end{eqnarray}
 with $F_{2}(1)=1/12$ and $x_{i}=M^2_{i}/m^2_{\eta^\pm}$. Note here that $B_{\alpha\beta}$ does not depend on the phases $\psi_{1,2}$ associated with heavy Majorana neutrino mass matrix.
From Eq.~(\ref{LFV}) the LFV branching ratios can be simplified as
\begin{eqnarray}
  {\rm Br}(\ell_{\alpha}\rightarrow \ell_{\beta}\gamma)&\simeq&8.0\times10^{5}r_{\alpha\beta}\frac{|B_{\alpha\beta}|^{2}}{m^{4}_{\eta^{\pm}}}{\rm GeV}^{4}~,
 \label{LFV1}
 \end{eqnarray}
where $r_{\mu e}=1.0, r_{\tau e}=0.1784$ and $r_{\tau\mu}=0.1736$, which indicates that the LFV branching ratios depends on $m_{\eta^{\pm}}$ and on the neutrino parameters in a flavor dependent manner.

In the following we consider the case, $M_{3}\simeq m_{\eta^{\pm}}\ll M_{2}<M_{1}$, as shown in section-III. Taking the case $M_{1}\simeq2M,~M_{2}\simeq M$ and $M_{3}\simeq0.01M$ into account, the function $F_{2}(x_{i})$ can have the values  1/12 $(x_{3}=1)$,  $3.3\times10^{-5}(~x_{2}=10^{4})$ and $8.3\times10^{-6}(~x_{1}=4\times10^{4})$, for $N_{1}, N_{2}$ and $N_{3}$, respectively, which indicates only the lightest heavy Majorana neutrino among the heavy neutrinos can contribute the branching ratio of LFV. Then, the expressions of $|B_{\alpha\beta}|^{2}$ relevant for $\tau\rightarrow\mu\gamma$, $\tau\rightarrow e\gamma$ and $\mu\rightarrow e\gamma$, respectively, are approximately given as
\begin{eqnarray}
  |B_{\tau\mu}|^{2}= |y_{\nu}|^{4}F^{2}_{2}(x_{3}) Q_{\tau\mu}~,\qquad\qquad
  |B_{\tau e}|^{2}\simeq |B_{\mu e}|^{2}= |y_{\nu}|^{4}F^{2}_{2}(x_{3}) Q_{\mu e}~,
 \label{Bab1}
 \end{eqnarray}
where the quantities $Q_{\tau\mu}$ and $Q_{\mu e}$ are simply expressed up to order of $\lambda^{2}$ as
\begin{eqnarray}
  Q_{\tau\mu}&\simeq& \frac{1}{4}-\frac{y_{1}\cos\rho_{1}+y_{2}\cos\rho_{2}}{2}-\frac{\lambda}{2\sqrt{3}}\left(y_{2}\sin(\rho_{2}-\phi^{\ell}_{3})-y_{1}\sin(\rho_{1}-\phi^{\ell}_{3})\right)-\frac{\lambda^{2}}{4}~,\nonumber\\
  Q_{\mu e}&\simeq& \frac{(y_{1}-y_{2})^{2}}{12}+\frac{y_{1}y_{2}}{3}\sin^{2}\left(\frac{\rho_{1}-\rho_{2}}{2}\right)\nonumber\\
&&+\frac{\lambda}{2\sqrt{3}}\left(y_{2}\sin(\rho_{2}-\phi^{\ell}_{3})-y_{1}\sin(\rho_{1}-\phi^{\ell}_{3})\right)+\frac{\lambda^{2}}{4}~.
 \label{Bab1}
 \end{eqnarray}
From the constraint of the branching ratio of $\tau\to\mu\gamma$ in Eq.~(\ref{expLFV}) we can obtain a lower bound of DM mass by using Eqs.~(\ref{coupling},\ref{LFV1},\ref{Bab1}):
 \begin{eqnarray}
  66~{\rm GeV}\lesssim M_{3}\lesssim83~{\rm TeV}~,
 \end{eqnarray}
where the upper bound is from Eq.~(\ref{upper}). Using Eq.~(\ref{expLFV}), the branching ratio Eq.~(\ref{LFV1}) can be expressed in terms of $Q_{\mu e}$, $Q_{\tau\mu}$ and $x_{\eta}$ as
 \begin{eqnarray}
  Q_{\mu e}\leq x^{2}_{\eta}\times1.26\times10^{-10}~,\qquad Q_{\tau\mu}\leq x^{2}_{\eta}\times1.33\times10^{-5}~,
 \end{eqnarray}
indicating $Q_{\mu e}$ is much more stringent than $Q_{\tau\mu}$ in sensitivity.
\begin{figure}[t]
\begin{center}
\includegraphics*[width=0.5\textwidth]{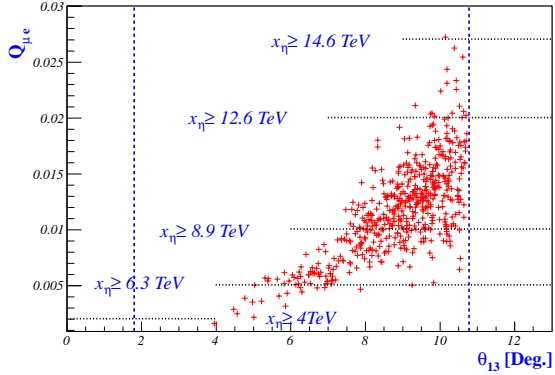}
\caption{The $Q_{\mu e}\lesssim x^{2}_{\eta}\times1.26\times10^{-10}$ sensitivity to $\theta_{13}$. Here the vertical dotted lines represent the upper and lower bounds in $3\sigma$ of Eq.~(\ref{expnu})}
\label{theta13-qume}
\end{center}
\end{figure}
In Fig.~\ref{theta13-qume}, the allowed region between $\theta_{13}$ and $Q_{\mu e}$ is described. It tells us that a rather large $\theta_{13}>9^\circ$ is predicted for 12.6 TeV $\le M_3\le$ 14.6 TeV.
For $M_3<$ 6.3 TeV, on the other hand, a rather smaller $\theta_{13}< 8^\circ$ is predicted.
It might be able to detected by direct or indirect DM searches, though the mass sale is rather high.

Finally we mention the detectability at the LHC.
We find that an inert charged Higgs, which decays into a muon and tauon (or a anti-muon and a anti-tauon) universally with a large missing energy
in the limit of leading order, where the missing energy is carried away by our DM.
It suggests that the clean signal will be obtained at the LHC.

\section{Conclusion}
We have analyzed the lepton masses and the mixings based on $A_4$ flavor symmetry. Especially, we obtained a lower bound of $\theta_{13}\gtrsim3.5^{\circ}$ in hierarchical normal mass ordering. And, we found that the large $\theta_{13}$ can be correlated with the Majorana type of DM, and its mass be ${\cal O}$(1-10) TeV,
as a result of analyzing the WMAP and lepton flavor violation such as the $\mu\rightarrow e\gamma$ experiment. Moreover, we could predict that a rather lager
$\theta_{13}$; $9^\circ<\theta_{13}$ for 12.6 TeV $\le M_3\le$ 14.6 TeV, and a rather smaller $\theta_{13}$; $\theta_{13}< 8^\circ$, for $M_3<$ 6.3 TeV. It implies that it might be able to detetected by direct or indirect DM searches as well as LHC at a rather high energy scale.

\appendix
\section{Higgs Potential and vacuum alignment}
\begin{figure}[h]
  \epsfig{figure=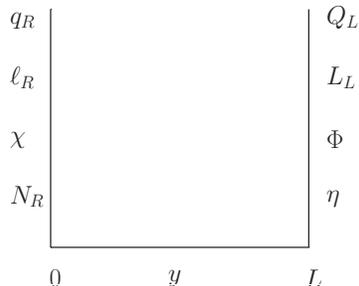,width=5cm,angle=0}
 \caption{\label{fig:exd}
  Fifth dimension and locations of scalar and fermion fields.}
\end{figure}

Since it is nontrivial to ensure that the different vacuum alignments of
$\langle\varphi^{0}\rangle=(\upsilon,\upsilon,\upsilon)$ and $\langle\chi\rangle=(\upsilon_{\chi},0,0)$
in Eq.~(\ref{subgroup}) are preserved, we shall briefly discuss these vacuum alignments.
There is a generic way to prohibit the problematic interaction terms by physically separating the
fields $\chi$ and $(\Phi,\eta)$.
Here we solve the vacuum alignment problem by extending the model with a spacial extra dimension
$y$~\cite{Altarelli:2005yp}. We assume that each field lives on the 4D brane either at $y = 0$ or at
$y = L$, as shown in Fig.~\ref{fig:exd}. The heavy neutrino masses arise from local operators at $y=0$,
while the charged fermion masses and the neutrino Yukawa interactions are realized by non-local effects
involving both branes.
A detailed explanation of this possibility is beyond the scope of this paper.

\vspace{0.5cm}
Then, the most general renormalizable scalar potential of $\Phi, \eta$ and $\chi$, invariant
under $SU(2)_{L}\times U(1)_{Y}\times A_{4}\times Z_{2}$, is given by
 \begin{eqnarray}
V_{y=L} &=& \mu^{2}_{\Phi}(\Phi^{\dag}\Phi)_{\mathbf{1}}
 +\lambda^{\Phi}_{1}(\Phi^{\dag}\Phi)_{\mathbf{1}}(\Phi^{\dag}\Phi)_{\mathbf{1}}
 +\lambda^{\Phi}_{2}(\Phi^{\dag}\Phi)_{\mathbf{1^\prime}}(\Phi^{\dag}\Phi)_{\mathbf{1^{\prime\prime}}}
 +\lambda^{\Phi}_{3}(\Phi^{\dag}\Phi)_{\mathbf{3}_{s}}(\Phi^{\dag}\Phi)_{\mathbf{3}_{s}}\nonumber\\
  &+&\lambda^{\Phi}_{4}(\Phi^{\dag}\Phi)_{\mathbf{3}_{a}}(\Phi^{\dag}\Phi)_{\mathbf{3}_{a}}
  +\lambda^{\Phi}_{5}(\Phi^{\dag}\Phi)_{\mathbf{3}_{s}}(\Phi^{\dag}\Phi)_{\mathbf{3}_{a}}
  +\mu^{2}_{\eta}(\eta^{\dag}\eta)+\lambda^{\eta}(\eta^{\dag}\eta)^{2}\nonumber\\
  &+&\lambda^{\Phi\eta}_{1}(\Phi^{\dag}\Phi)_{\mathbf{1}}(\eta^{\dag}\eta)
  +\lambda^{\Phi\eta}_{2}(\Phi^{\dag}\eta)(\eta^{\dag}\Phi)
  +\lambda^{\Phi\eta}_{3}(\Phi^{\dag}\eta)(\Phi^{\dag}\eta)
  +\lambda^{\Phi\eta\ast}_{3}(\eta^{\dag}\Phi)(\eta^{\dag}\Phi) ~,
\label{potential1}\\
V_{y=0}  &=&\mu^{2}_{\chi}(\chi\chi)_{\mathbf{1}}
 +\xi^{\chi}(\chi\chi\chi)_{\mathbf{1}}
 +\lambda^{\chi}_{1}(\chi\chi)_{\mathbf{1}}(\chi\chi)_{\mathbf{1}}
  +\lambda^{\chi}_{2}(\chi\chi)_{\mathbf{1}^\prime}(\chi\chi)_{\mathbf{1}^{\prime\prime}}
  +\lambda^{\chi}_{3}(\chi\chi)_{\mathbf{3}}(\chi\chi)_{\mathbf{3}}\nonumber\\
  &+&
\frac{\kappa_1}{\Lambda}(\chi\chi)_{\mathbf{1}}(\chi\chi)_{\mathbf{3}}\chi
+\frac{\kappa_2}{\Lambda}(\chi\chi)_{\mathbf{1^\prime}}(\chi\chi)_{\mathbf{3}}\chi
+\frac{\kappa_3}{\Lambda}(\chi\chi)_{\mathbf{1^{\prime\prime}}}(\chi\chi)_{\mathbf{3}}\chi
+\frac{\kappa_4}{\Lambda}(\chi\chi)_{\mathbf{3}}(\chi\chi)_{\mathbf{3}}\chi
,
\label{potential2}
\end{eqnarray}
where $\mu_{\Phi},\mu_{\eta},\mu_{\chi}$ and $\xi^{\chi}$ are of the mass dimension 1, while
$\lambda^{\Phi}_{1,...,5}$, $\lambda^{\eta}$, $\lambda^{\chi}_{1,...,3}$,
$\lambda^{\Phi\eta}_{1,...,3}$ and $\kappa_{1,...,4}$ are all dimensionless. Notice also here that there are no five dimensional terms in $V_{y=L}$ and $\kappa_2=\kappa_3=0$.
Let us set $\varphi^0_i\equiv v + \varphi_{Ri} + i~\varphi_{Ii}$ $(i=1,2,3)$
and $\eta^0\equiv \eta^{0}_{R}+i~\eta^{0}_{R}$. Then masses of $\Phi_{i}$, $\eta$, and $\chi_i$ are written as follows:
\begin{eqnarray}
&& m^2_R \equiv m^2({\rm Re}(\eta)) = \frac12\mu^2_{\eta}+ \frac32v^2(\lambda^{\Phi\eta}_{1}+\lambda^{\Phi\eta}_{2}+2\lambda^{\Phi\eta}_{3})~,\\
&& m^2_I \equiv m^2({\rm Im}(\eta)) =  \frac12\mu^2_{\eta}+ \frac32v^2(\lambda^{\Phi\eta}_{1}+\lambda^{\Phi\eta}_{2}-2\lambda^{\Phi\eta}_{3})~,\\
&& m^{2}_{\eta^\pm} =  \frac12\mu^2_{\eta}+ \frac32v^2\lambda^{\Phi\eta}_{1}~,\\
&& m^2_{ {\rm Re}(\varphi)_i}
=
{\left(\begin{array}{ccc}
  2(\lambda^{\Phi}_1+\lambda^{\Phi}_2) &  2\lambda^{\Phi}_1-\lambda^{\Phi}_2+4\lambda^{\Phi}_3 & 2\lambda^{\Phi}_1-\lambda^{\Phi}_2+4\lambda^{\Phi}_3 \\
  2\lambda^{\Phi}_1-\lambda^{\Phi}_2+4\lambda^{\Phi}_3 &  2(\lambda^{\Phi}_1+\lambda^{\Phi}_2) & 2\lambda^{\Phi}_1-\lambda^{\Phi}_2+4\lambda^{\Phi}_3  \\
  2\lambda^{\Phi}_1-\lambda^{\Phi}_2+4\lambda^{\Phi}_3 & 2\lambda^{\Phi}_1-\lambda^{\Phi}_2+4\lambda^{\Phi}_3 &  2(\lambda^{\Phi}_1+\lambda^{\Phi}_2) \end{array}\right)} v^2,\\
  && m^2_{ {\rm Im}(\varphi)_i}
=
2{\left(\begin{array}{ccc}
  -2(\lambda^{\Phi}_3+\lambda^{\Phi}_4) &  (\lambda^{\Phi}_3+\lambda^{\Phi}_4)  & (\lambda^{\Phi}_3+\lambda^{\Phi}_4) \\
 (\lambda^{\Phi}_3+\lambda^{\Phi}_4) &  -2(\lambda^{\Phi}_3+\lambda^{\Phi}_4)  & (\lambda^{\Phi}_3+\lambda^{\Phi}_4) \\
 (\lambda^{\Phi}_3+\lambda^{\Phi}_4) & (\lambda^{\Phi}_3+\lambda^{\Phi}_4) &  -2(\lambda^{\Phi}_3+\lambda^{\Phi}_4)  \end{array}\right)} v^2,\\
   && m^2_{ (\varphi^\pm)_i}
=
4{\left(\begin{array}{ccc}
  -2\lambda^{\Phi}_3 &   \lambda^{\Phi}_3  & \lambda^{\Phi}_3 \\
 \lambda^{\Phi}_3 &   -2\lambda^{\Phi}_3  & \lambda^{\Phi}_3 \\
 \lambda^{\Phi}_3& \lambda^{\Phi}_3 &  -2\lambda^{\Phi}_3  \end{array}\right)} v^2,\\
    &&
    m^2_{ [{\rm Re}(\varphi)-{\rm Im}(\varphi)]_i}
=
4{\left(\begin{array}{ccc}
0 &   -\lambda^{\Phi}_5  & \lambda^{\Phi}_5 \\
 \lambda^{\Phi}_5 &   0  & -\lambda^{\Phi}_5 \\
 -\lambda^{\Phi}_5 & \lambda^{\Phi}_5 & 0  \end{array}\right)} v^2,\\
    && m^2_{ (\chi)_i}
=
diag\left(2(\lambda^\chi_1+\lambda^\chi_2)~,~\lambda^\chi_1-\frac12\lambda^\chi_2+2\lambda^\chi_3~,~\lambda^\chi_1-\frac12\lambda^\chi_2+2\lambda^\chi_3   \right),
\end{eqnarray}
where we assume $\lambda^{\Phi\eta}_{3}$ is real to forbid the mixing term between  ${\rm Re}(\eta)$ and  ${\rm Im}(\eta)$, and also $\chi$ is real. Here we use stable conditions; $\mu^2_{\Phi}=-(6\lambda^{\Phi}_1+8\lambda^{\Phi}_3)v^2$ and $\mu^2_{\chi}=-2(\lambda^{\chi}_1+\lambda^{\chi}_2)v^2_\chi$.


\end{document}